\PassOptionsToPackage{unicode}{hyperref}
\PassOptionsToPackage{hyphens}{url}
\documentclass[
  11pt]{article}
\usepackage{xcolor}
\usepackage[margin=1in]{geometry}
\usepackage{amsmath,amssymb}
\usepackage{iftex}
\ifPDFTeX
  \usepackage[T1]{fontenc}
  \usepackage[utf8]{inputenc}
  \usepackage{textcomp} 
\else 
  \usepackage{unicode-math} 
  \defaultfontfeatures{Scale=MatchLowercase}
  \defaultfontfeatures[\rmfamily]{Ligatures=TeX,Scale=1}
\fi
\usepackage{lmodern}
\ifPDFTeX\else
\fi
\IfFileExists{upquote.sty}{\usepackage{upquote}}{}
\IfFileExists{microtype.sty}{
  \usepackage[]{microtype}
  \UseMicrotypeSet[protrusion]{basicmath} 
}{}
\makeatletter
\@ifundefined{KOMAClassName}{
  \IfFileExists{parskip.sty}{%
    \usepackage{parskip}
  }{
    \setlength{\parindent}{0pt}
    \setlength{\parskip}{6pt plus 2pt minus 1pt}}
}{
  \KOMAoptions{parskip=half}}
\makeatother
\usepackage{color}
\usepackage{fancyvrb}

\DefineVerbatimEnvironment{Highlighting}{Verbatim}{commandchars=\\\{\}}
\usepackage{framed}
\definecolor{shadecolor}{RGB}{248,248,248}
\newenvironment{Shaded}{\begin{snugshade}}{\end{snugshade}}

\newcommand{\AttributeTok}[1]{\textcolor[rgb]{0.13,0.29,0.53}{#1}}

\newcommand{\CommentTok}[1]{\textcolor[rgb]{0.56,0.35,0.01}{\textit{#1}}}

\newcommand{\ControlFlowTok}[1]{\textcolor[rgb]{0.13,0.29,0.53}{\textbf{#1}}}

\newcommand{\DecValTok}[1]{\textcolor[rgb]{0.00,0.00,0.81}{#1}}

\newcommand{\FloatTok}[1]{\textcolor[rgb]{0.00,0.00,0.81}{#1}}
\newcommand{\FunctionTok}[1]{\textcolor[rgb]{0.13,0.29,0.53}{\textbf{#1}}}

\newcommand{\NormalTok}[1]{#1}

\newcommand{\OtherTok}[1]{\textcolor[rgb]{0.56,0.35,0.01}{#1}}

\newcommand{\SpecialCharTok}[1]{\textcolor[rgb]{0.81,0.36,0.00}{\textbf{#1}}}

\newcommand{\StringTok}[1]{\textcolor[rgb]{0.31,0.60,0.02}{#1}}

\usepackage{longtable,booktabs,array}
\usepackage{calc} 
\usepackage{etoolbox}
\makeatletter
\patchcmd\longtable{\par}{\if@noskipsec\mbox{}\fi\par}{}{}
\makeatother
\IfFileExists{footnotehyper.sty}{\usepackage{footnotehyper}}{\usepackage{footnote}}
\makesavenoteenv{longtable}
\usepackage{graphicx}
\makeatletter
\newsavebox\pandoc@box
\newcommand*\pandocbounded[1]{
  \sbox\pandoc@box{#1}%
  \Gscale@div\@tempa{\textheight}{\dimexpr\ht\pandoc@box+\dp\pandoc@box\relax}%
  \Gscale@div\@tempb{\linewidth}{\wd\pandoc@box}%
  \ifdim\@tempb\p@<\@tempa\p@\let\@tempa\@tempb\fi
  \ifdim\@tempa\p@<\p@\scalebox{\@tempa}{\usebox\pandoc@box}%
  \else\usebox{\pandoc@box}%
  \fi%
}
\def\fps@figure{htbp}
\makeatother
\providecommand{\tightlist}{%
  \setlength{\itemsep}{0pt}\setlength{\parskip}{0pt}}
\usepackage[]{natbib}
\usepackage{lscape}
\usepackage{float}
\usepackage{multirow}
\usepackage{authblk}
\newcolumntype{L}[1]{>{\raggedright\arraybackslash}p{#1}}

\renewcommand{\and}{\qquad}
\AtBeginDocument{
  \let\maketitle\relax
  
  \begin{center}
    {\LARGE\bfseries fastml: Guarded Resampling Workflows for Safer Automated Machine Learning in R\par}
    \vspace{0.8em}
    {\large Selcuk Korkmaz\textsuperscript{1}\footnote[1]{Corresponding author: \href{mailto:selcukorkmaz@gmail.com}{selcukorkmaz@gmail.com}, ORCID: 0000-0003-4632-6850}, \quad Dincer Goksuluk\textsuperscript{2}\footnote[2]{ORCID: 0000-0002-2752-7668}, \quad Eda Karaismailoglu\textsuperscript{3}\footnote[3]{ORCID: 0000-0003-3085-7809}\par}
    \vspace{0.5em}
    {\small \textsuperscript{1}Faculty of Medicine, Department of Biostatistics, Trakya University, Edirne 22030, T\"{u}rkiye\\
    \textsuperscript{2}Faculty of Medicine, Department of Biostatistics, Sakarya University, Sakarya 54290, T\"{u}rkiye\\
    \textsuperscript{3}Faculty of Medicine, Department of Biostatistics, University of Health Sciences, Ankara 06010, T\"{u}rkiye\par}
    \vspace{1em}
  \end{center}
  
  \setcounter{footnote}{0}
}

\AtBeginEnvironment{landscape}{\setlength{\linewidth}{\textheight}\setlength{\textwidth}{\textheight}}
\makeatletter
\providecommand{\pandocbounded}[1]{#1}
\makeatother
\usepackage{bookmark}
\IfFileExists{xurl.sty}{\usepackage{xurl}}{} 
\hypersetup{
  pdfauthor={Selcuk Korkmaz, Dincer Goksuluk, Eda Karaismailoglu},
  pdfkeywords={machine learning, preprocessing leakage, cross-validation, resampling, survival analysis},
  hidelinks,
  pdfcreator={LaTeX via pandoc}}

\title{\textbf{fastml: Guarded Resampling Workflows for Safer Automated Machine Learning in R}}
\author{Selcuk Korkmaz, Dincer Goksuluk, Eda Karaismailoglu}
\date{}

\begin{document}
\maketitle
\begin{abstract}
Preprocessing leakage arises when scaling, imputation, or other data-dependent transformations are estimated before resampling, inflating apparent performance while remaining hard to detect. We present fastml, an R package that provides a single-call interface for leakage-aware machine learning through guarded resampling, where preprocessing is re-estimated inside each resample and applied to the corresponding assessment data. The package supports grouped and time-ordered resampling, blocks high-risk configurations, audits recipes for external dependencies, and includes sandboxed execution and integrated model explanation. We evaluate fastml with a Monte Carlo simulation contrasting global and fold-local normalization, a usability comparison with tidymodels under matched specifications, and survival benchmarks across datasets of different sizes. The simulation demonstrates that global preprocessing substantially inflates apparent performance relative to guarded resampling. fastml matched held-out performance obtained with tidymodels while reducing workflow orchestration, and it supported consistent benchmarking of multiple survival model classes through a unified interface.
\end{abstract}

\vspace{0.5em}


\vspace{1em}

\section{Introduction}\label{sec:introduction}

Over the past decade, machine learning workflows in R have evolved from monolithic interfaces, such as \textbf{caret} \citep{caret_kuhn:2008aa}, toward modular ecosystems including \textbf{tidymodels} \citep{tidymodels_Kuhn:2020aa} and \textbf{mlr3} \citep{mlr3_Lang:2019aa}. These frameworks decompose modeling into explicit steps for resampling, preprocessing, model specification, and evaluation, providing flexibility and methodological transparency. However, this modularity also shifts responsibility to the user to correctly assemble and coordinate multiple components into a valid pipeline. This shift increases the risk of methodological error by expanding the number of decisions that must be coordinated correctly. Users must ensure that preprocessing, resampling, and model fitting are applied in the proper order and with strict separation between training and evaluation data. In complex workflows, particularly those involving grouped or nested resampling, even experienced practitioners may inadvertently introduce errors that are not detected by standard software checks.

Among these errors, data leakage remains one of the most consequential. In particular, preprocessing leakage occurs when transformations such as scaling, imputation, or feature construction are estimated using information that should be unavailable at evaluation time, for example by learning transformations on the full dataset prior to resampling. This violates training-test isolation and can substantially inflate performance estimates while remaining difficult to detect, because models may still appear well-validated under standard cross-validation \citep{Hastie:2009_ESL_book}. Recent studies report that leakage and related evaluation flaws are widespread in applied machine learning and contribute to irreproducible and overly optimistic results across multiple scientific domains \citep{Kapoor:2023aa, Rosenblatt:2024aa, Tampu:2022aa}.

Modern R frameworks provide tools that can prevent preprocessing leakage, most notably by coupling transformations to resampling-aware workflows. However, these safeguards are typically optional rather than enforced. Correct usage requires users to explicitly define recipes or workflows, ensure that preprocessing is trained within each resampling split, and avoid global data-dependent transformations prior to model evaluation. When these conditions are not met, incorrect pipelines can run without warnings while yielding biased estimates. As a result, methodological validity often depends more on user discipline than on guarantees provided by the software. For example, tidymodels' \texttt{fit\_resamples()} and \texttt{tune\_grid()} correctly re-estimate preprocessing within each fold when a recipe is embedded in a workflow; however, users must assemble this workflow correctly, and nothing prevents them from applying global preprocessing before invoking these functions.

The \textbf{fastml} package \citep{Korkmaz:2025_fastml_Rpkg} addresses this gap through a safety-by-design interface that makes fold-local preprocessing the default execution path while retaining access to established modeling engines. A central mechanism is its guarded resampling path, in which preprocessing is re-estimated within each resample on the analysis split and the fold-trained transformations are then applied to the corresponding assessment split. \textbf{fastml} also includes checks intended to surface common configuration errors, including aborting a resampling run when a full-analysis (no-holdout) split is detected and rejecting user recipes that appear to depend on external data sources. When hyperparameter tuning is requested under this guarded path, tuning is performed inside the same resampling loop.

More broadly, \textbf{fastml} provides resampling options and diagnostics that target additional evaluation pitfalls. Grouped resampling can keep related observations together when repeated measures or clustered records exist. For time-ordered settings, blocked or rolling resampling relies on an explicit ordering variable and \textbf{fastml} warns when the ordering required by the chosen design is not provided. An optional audit mode can flag risky patterns in custom preprocessing, such as references to the global environment or code that reads data from files or writes outputs to files, which can create hidden dependencies and undermine reproducibility.

These mechanisms are best understood as leakage-risk mitigation under supported configurations rather than universal guarantees; users can still introduce leakage upstream, for example by supplying preprocessed inputs or externally constructed resamples. The contribution of \textbf{fastml} is therefore not a new learning algorithm, but an execution model that makes leakage-aware evaluation easier to carry out correctly in practice by packaging common patterns, defaults, and checks into a single interface. This paper describes the design and implementation of \textbf{fastml}, details how guarded resampling and related checks are realized when those execution paths are used, and demonstrates via simulation how preprocessing leakage can inflate performance estimates and how fold-local workflow fitting mitigates this failure mode under the evaluated configurations.

The remainder of this paper is organized as follows. Section 2 presents the guarded resampling architecture and its implementation, introduces a Monte Carlo simulation designed to quantify the impact of preprocessing leakage, describes the native survival analysis interface, outlines the software architecture and audit utilities, and compares \textbf{fastml} with existing frameworks. Section 3 reports the simulation results, followed by applied benchmarks and case studies in classification, regression, and survival analysis. Section 4 discusses limitations and future directions.

\section{Materials and methods}\label{materials-and-methods}

\subsection{Guarded resampling}\label{sec:guarded-resampling}

As introduced in the previous section, preprocessing leakage occurs when data-dependent transformation parameters are estimated on the full dataset before resampling, so that assessment-fold observations influence training-fold transformations \citep{Kapoor:2023aa, Kaufman:2012aa, Vabalas:2019aa}. This breaks training-test separation and tends to yield optimistic performance estimates, even for unsupervised transformations that never access outcome labels \citep{Hastie:2009_ESL_book, Kaufman:2012aa}.

\begin{figure}
\includegraphics[width=1\linewidth,alt={Two side-by-side workflow diagrams compare leaky and guarded cross-validation. The leaky workflow applies preprocessing before resampling, while the guarded workflow estimates preprocessing separately inside each fold before evaluation.}]{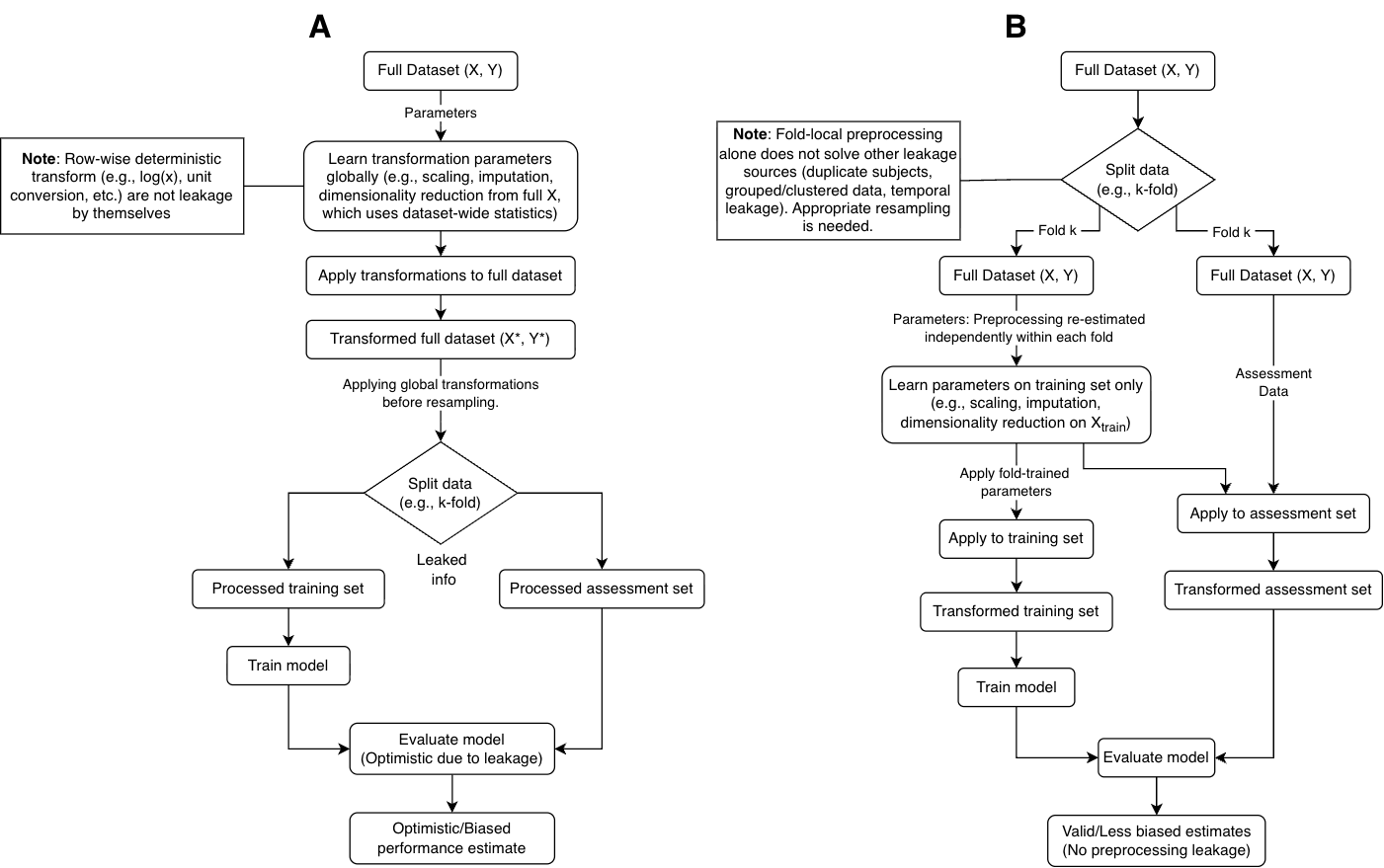} \caption{Illustration of a leaky cross-validation workflow caused by global preprocessing.}\label{fig:leakage-workflow}
\end{figure}

Although existing R frameworks can avoid this problem by embedding transformations within resampling-aware workflows, such isolation is optional, and leaky pipelines can execute without warnings. Figure \ref{fig:leakage-workflow} illustrates the distinction. In a leaky workflow, transformation parameters are estimated on the full dataset before resampling, so each fold's assessment data influences the transformation applied to that fold's training data. In a properly specified workflow, transformation parameters are estimated using only the training portion of each fold and then applied to that fold's assessment portion.

When \textbf{fastml} executes its guarded resampling path (i.e., when workflows are fitted through \texttt{fastml\_guarded\_resample\_fit}), it organizes preprocessing and model fitting fold-by-fold to reduce preprocessing-leakage risk, as introduced in the Introduction.

\subsubsection{The guarded architecture}\label{the-guarded-architecture}

Under guarded resampling, any data-dependent preprocessing is re-estimated within each resample. For each fold \(k\), the procedure is:

\begin{itemize}
\item
  \textbf{Generate split \(k\)}: Construct a resampling split and obtain an analysis (training) set \(D_{analysis}^{k}\) and an assessment (hold-out) set \(D_{assess}^{k}\).
\item
  \textbf{Fit fold-specific preprocessing}: Fit a fresh preprocessing specification \(R^{(k)}\) using only \(D_{analysis}^{(k)}\). This includes estimating any data-dependent transformation parameters (e.g., imputation values, scaling parameters, feature construction rules) from the fold's analysis data.
\item
  \textbf{Apply fold-trained preprocessing}: Apply \(R^{(k)}\) to \(D_{analysis}^{(k)}\) to produce the preprocessed training features, and apply the same fold-trained preprocessing \(R^{(k)}\) to \(D_{assess}^{(k)}\) to produce the corresponding preprocessed assessment features. The assessment set is not used when fitting \(R^{(k)}\); it is only transformed after \(R^{(k)}\) has been learned from the analysis data.
\item
  \textbf{Fit and evaluate within the fold}: Train the model on the preprocessed \(D_{analysis}^{(k)}\) and evaluate it on the transformed \(D_{assess}^{(k)}\), producing performance for fold \(k\).
\end{itemize}

Repeating these steps across folds ensures that, within the resampling process, data-dependent preprocessing parameters are estimated without access to the corresponding assessment data.

\subsubsection{Implementation in fastml}\label{implementation-in-fastml}

In \textbf{fastml}, the guarded resampling behavior is implemented in the resampling engine used when the package executes its guarded resampling path. The core implementation is \texttt{fastml\_guarded\_resample\_fit}, which iterates directly over \textbf{rsample} \citep{Frick:2025_rsample_Rpkg} split objects. For each split \(k\), the guarded resampling engine:

\begin{itemize}
\tightlist
\item
  Extracts the analysis (training) and assessment (hold-out) portions of the raw data for that split.
\item
  Fits the user-defined workflow, including preprocessing via \textbf{recipes} \citep{Kuhn:2025_recipes_Rpkg}, on the analysis portion of the split.
\item
  Applies the fitted preprocessing to the assessment portion and evaluates the trained model on that fold's assessment data.
\end{itemize}

This structure ensures that, within the guarded resampling loop, recipes are trained on the analysis portion of each split and then applied to the corresponding assessment portion for evaluation. The guard also performs a structural check to detect pathological resampling definitions. In particular, it detects ``full-analysis'' splits, cases where the analysis indices cover the full dataset and therefore no holdout remains, and stops with an error when such a configuration is encountered. This check is implemented by inspecting resample indices (the \texttt{in\_id} index sets), rather than by comparing full data objects directly. As noted in Section 1, upstream preprocessing or externally constructed resamples remain outside the scope of these guards.

In addition to the resampling structure, \textbf{fastml} includes a recipe-scanning guard that targets a narrow class of high-impact problems: it scans recipe steps for references to external environments or global objects and flags steps that appear to embed external data objects. Recipes failing these checks are rejected before model training begins. The full scope and limitations of this guard are detailed in the Security and audit utilities section.

Taken together, \textbf{fastml}'s guarded resampling functionality reduces leakage risk by (i) fitting workflows within each resample when executed through \texttt{fastml\_guarded\_resample\_fit}, (ii) detecting full-analysis resampling failures by checking \texttt{in\_id} indices, and (iii) blocking recipe steps that appear to reference external environments or embed external data objects.

Beyond preprocessing leakage, \textbf{fastml} also supports resampling designs that reduce common evaluation artifacts. Grouped cross-validation keeps related rows together to reduce group-based leakage \citep{Kuhn:2013_APM}, and blocked/rolling resampling modes surface warnings that correct ordering is required to avoid training on future information \citep{Hyndman:2018_forecasting}.

\subsubsection{Resampling design for grouped and time-ordered data}\label{resampling-design-for-grouped-and-time-ordered-data}

Fold-local preprocessing addresses a specific leakage mechanism: estimating data-dependent transformation parameters using information that should be unavailable at evaluation time. However, valid evaluation can still fail when the resampling design does not match the dependence structure of the data. Two common cases are grouped (clustered) data and time-ordered data.

In grouped or clustered settings, optimistic evaluation can occur even when preprocessing is strictly fold-local. If correlated records from the same entity (e.g., repeated measurements from a patient, multiple samples from the same subject, or clustered observations from the same unit) are split across folds, models can exploit within-entity similarity and appear to generalize better than they truly do \citep{Roberts:2017aa}. To reduce this risk, \textbf{fastml} supports grouped resampling designs in which all rows sharing a grouping identifier are assigned to the same split, using \texttt{group\_cols} under the corresponding grouped resampling method. This can mitigate group-based leakage when the grouping variable captures the relevant dependence structure.

In time-ordered settings, leakage can arise when the resampling scheme allows training data to include observations occurring after the assessment period, producing ``future-to-past'' contamination. This can occur without any explicit preprocessing mistake, purely because the split violates the temporal direction implied by the prediction task \citep{Bergmeir:2012aa}. \textbf{fastml} supports blocked and rolling resampling modes intended for such settings, which rely on an explicit ordering variable (e.g., \texttt{block\_col}) and associated block/window parameters. Because correct ordering is a prerequisite for valid time-aware evaluation, \textbf{fastml} warns when the ordering required by the chosen resampling design is not provided. As with grouped designs, these options mitigate a common leakage route under appropriate configuration \citep{Hyndman:2018_forecasting}.

These resampling designs therefore complement fold-local preprocessing: guarded resampling primarily targets preprocessing leakage within resampling, while grouped and time-aware resampling options help reduce leakage that arises from dependence structures and temporal ordering.

\subsubsection{Failure-mode demonstration: impact of improper batch correction}\label{failure-mode-demonstration-impact-of-improper-batch-correction}

A central motivation for guarded resampling is that preprocessing leakage can arise even in workflows that otherwise appear methodologically careful. In multi-site or multi-center studies, a particularly common failure mode is improper batch correction, where site-specific normalization is performed globally prior to resampling. Although often intended to reduce heterogeneity, such preprocessing can incorporate information from held-out groups and inflate apparent performance estimates \citep{Luo:2010aa}.

To demonstrate this failure mode and to illustrate the consequences of global preprocessing under grouped validation, we conducted a Monte Carlo simulation contrasting improper global batch correction with fold-specific preprocessing embedded in a resampling workflow. This simulation is presented as \emph{external experimental evidence in the paper}, not as a built-in demonstration performed automatically by the package. The extent to which fastml's guarded resampling path is used depends on internal execution logic and the configuration of the modeling call. The complete reproducibility code for this simulation is provided in the supplementary file \texttt{failure\_mode\_simulation.R}.

\textbf{Simulation setting.}

We simulate a binary classification problem with a latent signal shared across sites and a strong site-specific batch effect in the observed predictor. For observation \(i\) belonging to site \(s(i)\), we generate a latent signal \(z_i \sim N(0,1)\), a site offset \(b_s \sim N(5,5)\), and an observed predictor \(x_i=z_i+b_s(i)\). The outcome depends only on the latent signal, not on site membership: \(p_i = logit^{-1}(2z_i)\), and \(y_i \sim \text{Bernoulli}(p_i)\). Evaluation uses grouped cross-validation, holding out one entire site per fold (leave-one-site-out), and the simulation is repeated over multiple independent runs to assess stability.

We compare two workflows:

\begin{itemize}
\item
  Leaky workflow (global site-wise standardization before resampling). The predictor is standardized within each site using the full dataset before folds are created. This induces leakage because observations that later appear in assessment folds contribute to the site-specific scaling parameters applied in training folds, partially aligning held-out site distributions with the training data and inflating AUC.
\item
  Guarded workflow (fold-specific preprocessing inside resampling via fastml). Preprocessing is defined in a recipe and is re-estimated on the analysis portion of each fold, then applied to the corresponding assessment portion. The site identifier is assigned an ID role to prevent it from being used as a predictor.
\end{itemize}

\begin{Shaded}
\begin{Highlighting}[]
\FunctionTok{library}\NormalTok{(dplyr)}
\FunctionTok{library}\NormalTok{(rsample)}
\FunctionTok{library}\NormalTok{(recipes)}
\FunctionTok{library}\NormalTok{(parsnip)}
\FunctionTok{library}\NormalTok{(workflows)}
\FunctionTok{library}\NormalTok{(yardstick)}
\FunctionTok{library}\NormalTok{(purrr)}
\FunctionTok{library}\NormalTok{(fastml)}
\FunctionTok{library}\NormalTok{(ggplot2)}
\FunctionTok{library}\NormalTok{(tidyr)}
\FunctionTok{library}\NormalTok{(patchwork)}

\FunctionTok{set.seed}\NormalTok{(}\DecValTok{2025}\NormalTok{)}

\NormalTok{n\_sims }\OtherTok{\textless{}{-}} \DecValTok{100}
\NormalTok{n\_sites }\OtherTok{\textless{}{-}} \DecValTok{10}
\NormalTok{n\_per\_site }\OtherTok{\textless{}{-}} \DecValTok{100}

\NormalTok{run\_iteration }\OtherTok{\textless{}{-}} \ControlFlowTok{function}\NormalTok{(seed) \{}
  \FunctionTok{set.seed}\NormalTok{(seed)}

\NormalTok{  total\_n }\OtherTok{\textless{}{-}}\NormalTok{ n\_sites }\SpecialCharTok{*}\NormalTok{ n\_per\_site}
\NormalTok{  sites }\OtherTok{\textless{}{-}} \FunctionTok{rep}\NormalTok{(}\FunctionTok{paste0}\NormalTok{(}\StringTok{"S"}\NormalTok{, }\DecValTok{1}\SpecialCharTok{:}\NormalTok{n\_sites), }\AttributeTok{each =}\NormalTok{ n\_per\_site)}

\NormalTok{  z }\OtherTok{\textless{}{-}} \FunctionTok{rnorm}\NormalTok{(total\_n, }\AttributeTok{mean =} \DecValTok{0}\NormalTok{, }\AttributeTok{sd =} \DecValTok{1}\NormalTok{)}
\NormalTok{  site\_offsets }\OtherTok{\textless{}{-}} \FunctionTok{rnorm}\NormalTok{(n\_sites, }\AttributeTok{mean =} \DecValTok{5}\NormalTok{, }\AttributeTok{sd =} \DecValTok{5}\NormalTok{)}
\NormalTok{  x\_observed }\OtherTok{\textless{}{-}}\NormalTok{ z }\SpecialCharTok{+} \FunctionTok{rep}\NormalTok{(site\_offsets, }\AttributeTok{each =}\NormalTok{ n\_per\_site)}

\NormalTok{  prob }\OtherTok{\textless{}{-}} \DecValTok{1} \SpecialCharTok{/}\NormalTok{ (}\DecValTok{1} \SpecialCharTok{+} \FunctionTok{exp}\NormalTok{(}\SpecialCharTok{{-}}\NormalTok{(z }\SpecialCharTok{*} \DecValTok{2}\NormalTok{)))}
\NormalTok{  y }\OtherTok{\textless{}{-}} \FunctionTok{factor}\NormalTok{(}\FunctionTok{ifelse}\NormalTok{(}\FunctionTok{runif}\NormalTok{(total\_n) }\SpecialCharTok{\textless{}}\NormalTok{ prob, }\StringTok{"Case"}\NormalTok{, }\StringTok{"Control"}\NormalTok{),}
    \AttributeTok{levels =} \FunctionTok{c}\NormalTok{(}\StringTok{"Control"}\NormalTok{, }\StringTok{"Case"}\NormalTok{)}
\NormalTok{  )}

\NormalTok{  df }\OtherTok{\textless{}{-}} \FunctionTok{tibble}\NormalTok{(}
    \AttributeTok{site =} \FunctionTok{factor}\NormalTok{(sites),}
    \AttributeTok{outcome =}\NormalTok{ y,}
    \AttributeTok{x =}\NormalTok{ x\_observed}
\NormalTok{)}

  \CommentTok{\# Leaky workflow:}
  \CommentTok{\# Global site{-}wise standardization before resampling}
\NormalTok{  df\_leaky }\OtherTok{\textless{}{-}}\NormalTok{ df }\SpecialCharTok{\%\textgreater{}\%}
    \FunctionTok{group\_by}\NormalTok{(site) }\SpecialCharTok{\%\textgreater{}\%}
    \FunctionTok{mutate}\NormalTok{(}\AttributeTok{x\_scaled =} \FunctionTok{as.numeric}\NormalTok{(}\FunctionTok{scale}\NormalTok{(x))) }\SpecialCharTok{\%\textgreater{}\%}
    \FunctionTok{ungroup}\NormalTok{()}

\NormalTok{  folds\_leaky }\OtherTok{\textless{}{-}} \FunctionTok{group\_vfold\_cv}\NormalTok{(}
\NormalTok{    df\_leaky,}
    \AttributeTok{group =} \StringTok{"site"}\NormalTok{,}
    \AttributeTok{v =}\NormalTok{ n\_sites}
\NormalTok{  )}

\NormalTok{  auc\_leaky }\OtherTok{\textless{}{-}} \FunctionTok{map\_dbl}\NormalTok{(folds\_leaky}\SpecialCharTok{$}\NormalTok{splits, }\ControlFlowTok{function}\NormalTok{(spl) \{}
\NormalTok{    model }\OtherTok{\textless{}{-}} \FunctionTok{rand\_forest}\NormalTok{(}\AttributeTok{trees =} \DecValTok{50}\NormalTok{) }\SpecialCharTok{\%\textgreater{}\%}
      \FunctionTok{set\_mode}\NormalTok{(}\StringTok{"classification"}\NormalTok{) }\SpecialCharTok{\%\textgreater{}\%}
      \FunctionTok{set\_engine}\NormalTok{(}\StringTok{"ranger"}\NormalTok{) }\SpecialCharTok{\%\textgreater{}\%}
      \FunctionTok{fit}\NormalTok{(outcome }\SpecialCharTok{\textasciitilde{}}\NormalTok{ x\_scaled, }\AttributeTok{data =} \FunctionTok{analysis}\NormalTok{(spl))}

    \FunctionTok{predict}\NormalTok{(model, }\FunctionTok{assessment}\NormalTok{(spl), }\AttributeTok{type =} \StringTok{"prob"}\NormalTok{) }\SpecialCharTok{\%\textgreater{}\%}
      \FunctionTok{bind\_cols}\NormalTok{(}\FunctionTok{assessment}\NormalTok{(spl)) }\SpecialCharTok{\%\textgreater{}\%}
      \FunctionTok{roc\_auc}\NormalTok{(}\AttributeTok{truth =}\NormalTok{ outcome, .pred\_Case,}
              \AttributeTok{event\_level =} \StringTok{"second"}\NormalTok{) }\SpecialCharTok{\%\textgreater{}\%}
      \FunctionTok{pull}\NormalTok{(.estimate)}
\NormalTok{  \})}

\NormalTok{  grouped\_folds }\OtherTok{\textless{}{-}}\NormalTok{ rsample}\SpecialCharTok{::}\FunctionTok{group\_vfold\_cv}\NormalTok{(}
\NormalTok{    df, }\AttributeTok{group =}\NormalTok{ site, }\AttributeTok{v =}\NormalTok{ n\_sites}
\NormalTok{  )}

  \CommentTok{\# Guarded workflow:}
  \CommentTok{\# Fold{-}specific preprocessing (fastml)}
\NormalTok{  guarded\_model }\OtherTok{\textless{}{-}} \FunctionTok{fastml}\NormalTok{(}
    \AttributeTok{data =}\NormalTok{ df,}
    \AttributeTok{label =} \StringTok{"outcome"}\NormalTok{,}
    \AttributeTok{algorithms =} \StringTok{"rand\_forest"}\NormalTok{,}
    \AttributeTok{resamples =}\NormalTok{ grouped\_folds,}
    \AttributeTok{metric =} \StringTok{"roc\_auc"}\NormalTok{,}
    \AttributeTok{event\_class =} \StringTok{"second"}\NormalTok{,}
    \AttributeTok{recipe =} \FunctionTok{recipe}\NormalTok{(outcome }\SpecialCharTok{\textasciitilde{}}\NormalTok{ x }\SpecialCharTok{+}\NormalTok{ site, }\AttributeTok{data =}\NormalTok{ df) }\SpecialCharTok{\%\textgreater{}\%}
      \FunctionTok{update\_role}\NormalTok{(site, }\AttributeTok{new\_role =} \StringTok{"id"}\NormalTok{) }\SpecialCharTok{\%\textgreater{}\%}
      \FunctionTok{step\_normalize}\NormalTok{(x),}
    \AttributeTok{tune\_params =} \FunctionTok{list}\NormalTok{(}
      \AttributeTok{rand\_forest =} \FunctionTok{list}\NormalTok{(}\AttributeTok{ranger =} \FunctionTok{list}\NormalTok{(}\AttributeTok{trees =} \DecValTok{50}\NormalTok{))}
\NormalTok{    )}
\NormalTok{  )}
  \FunctionTok{tibble}\NormalTok{(}
    \AttributeTok{Leaky\_AUC =} \FunctionTok{mean}\NormalTok{(auc\_leaky),}
    \AttributeTok{Guarded\_AUC =}\NormalTok{ guarded\_model}\SpecialCharTok{$}\NormalTok{performance}\SpecialCharTok{$}\NormalTok{rand\_forest}\SpecialCharTok{$}\NormalTok{ranger }\SpecialCharTok{\%\textgreater{}\%}
      \FunctionTok{filter}\NormalTok{(.metric }\SpecialCharTok{==} \StringTok{"roc\_auc"}\NormalTok{) }\SpecialCharTok{\%\textgreater{}\%}
      \FunctionTok{pull}\NormalTok{(.estimate)}
\NormalTok{  )}
\NormalTok{\}}
\end{Highlighting}
\end{Shaded}

In this simulation, \texttt{tune\_params} is used to fix the number of trees at 50 rather than to define a tuning grid; when a single value is supplied for a parameter, fastml treats it as a fixed setting.

We repeated the simulation for \texttt{n\_sims} independent runs. In each run, we computed a leaky AUC (site-wise scaling performed globally before grouped CV) and a guarded AUC (fold-local preprocessing inside grouped resampling). We summarize the distribution of AUC values across runs and quantify inflation using the paired difference (Leaky minus Guarded) across simulation runs.

The results of this simulation are reported in the Simulation results section.

\subsection{Package interface}\label{sec:package-interface}

The central entry point is \texttt{fastml()}, a single function that coordinates data splitting, preprocessing, model fitting, evaluation, and (optionally) hyperparameter tuning. Its arguments are grouped by purpose:

\begin{itemize}
\tightlist
\item
  \textbf{Data}: \texttt{data} (unsplit), or \texttt{train\_data}/\texttt{test\_data} (pre-split); \texttt{label} (outcome column); \texttt{test\_size} (holdout fraction, default 0.2).
\item
  \textbf{Algorithms}: \texttt{algorithms} (character vector or \texttt{"all"}); \texttt{algorithm\_engines} (optional engine overrides); \texttt{task} (\texttt{"auto"}, \texttt{"classification"}, \texttt{"regression"}, or \texttt{"survival"}).
\item
  \textbf{Preprocessing}: \texttt{recipe} (user-supplied recipes object), \texttt{impute\_method}, \texttt{encode\_categoricals}, \texttt{scaling\_methods}, \texttt{balance\_method}.
\item
  \textbf{Resampling}: \texttt{resampling\_method} (e.g., \texttt{"cv"}, \texttt{"repeatedcv"}, \texttt{"boot"}, \texttt{"grouped\_cv"}, \texttt{"blocked\_cv"}, \texttt{"rolling\_origin"}, \texttt{"nested\_cv"}, \texttt{"validation\_split"}, \texttt{"none"}); \texttt{folds}, \texttt{repeats}, \texttt{group\_cols}, \texttt{block\_col}, \texttt{resamples} (user-supplied rsample object).
\item
  \textbf{Tuning}: \texttt{tune\_params} (named list of grids per algorithm--engine pair); \texttt{tuning\_strategy} (\texttt{"grid"} or \texttt{"bayes"}); \texttt{tuning\_complexity} (\texttt{"quick"}, \texttt{"balanced"}, \texttt{"thorough"}, \texttt{"exhaustive"}); \texttt{use\_default\_tuning}.
\item
  \textbf{Evaluation}: \texttt{metric} (primary metric); \texttt{event\_class}; \texttt{class\_threshold}; \texttt{bootstrap\_ci} (default \texttt{TRUE}); \texttt{bootstrap\_samples}.
\item
  \textbf{Execution}: \texttt{n\_cores} (parallel workers); \texttt{seed}; \texttt{verbose}; \texttt{audit\_mode}.
\end{itemize}

\textbf{Supported algorithms.} fastml exposes the following algorithm families, each dispatched to one or more engines:

\begin{itemize}
\tightlist
\item
  \emph{Classification} (15 families): \texttt{logistic\_reg}, \texttt{multinom\_reg}, \texttt{decision\_tree}, \texttt{C5\_rules}, \texttt{rand\_forest}, \texttt{xgboost}, \texttt{lightgbm}, \texttt{svm\_linear}, \texttt{svm\_rbf}, \texttt{nearest\_neighbor}, \texttt{naive\_Bayes}, \texttt{mlp}, \texttt{discrim\_linear}, \texttt{discrim\_quad}, \texttt{bag\_tree}.
\item
  \emph{Regression} (14 families): \texttt{linear\_reg}, \texttt{ridge\_reg}, \texttt{lasso\_reg}, \texttt{elastic\_net}, \texttt{decision\_tree}, \texttt{rand\_forest}, \texttt{xgboost}, \texttt{lightgbm}, \texttt{svm\_linear}, \texttt{svm\_rbf}, \texttt{nearest\_neighbor}, \texttt{mlp}, \texttt{pls}, \texttt{bayes\_glm}.
\item
  \emph{Survival} (12 families): \texttt{cox\_ph}, \texttt{penalized\_cox}, \texttt{stratified\_cox}, \texttt{time\_varying\_cox}, \texttt{survreg}, \texttt{parametric\_surv}, \texttt{royston\_parmar}, \texttt{piecewise\_exp}, \texttt{rand\_forest}, \texttt{rand\_forest\_survival}, \texttt{xgboost}, \texttt{xgboost\_aft}. (Note: \texttt{rand\_forest} and \texttt{rand\_forest\_survival} map to different engines -- ranger and aorsf, respectively.)
\end{itemize}

\textbf{Resampling methods.} Nine strategies are supported: \texttt{cv} (k-fold cross-validation, default for classification/regression), \texttt{repeatedcv}, \texttt{boot} (bootstrap), \texttt{grouped\_cv} (requires \texttt{group\_cols}), \texttt{blocked\_cv} (requires \texttt{block\_col}), \texttt{rolling\_origin} (requires \texttt{initial\_window} and \texttt{assess\_window}), \texttt{nested\_cv} (requires \texttt{outer\_folds}), \texttt{validation\_split}, and \texttt{none} (default for survival).

\textbf{Return value and key methods.} \texttt{fastml()} returns an S3 object of class \texttt{fastml} that aggregates fitted models, the preprocessing specification, evaluation outputs, and (optionally) the data partitions (see the fastml object table below for details). The object supports the following methods:

\begin{itemize}
\tightlist
\item
  \texttt{predict.fastml(object,\ newdata)} -- applies the stored preprocessor and fitted model to new data; supports \texttt{type} options including \texttt{"class"}, \texttt{"prob"}, \texttt{"numeric"}, and \texttt{"survival"}.
\item
  \texttt{summary.fastml(object)} -- produces a two-table report that separates model selection from final evaluation. Table 1 (Model Selection) reports cross-validation performance (mean and SD of the primary metric across folds) and identifies the best model; Table 2 (Final Evaluation) reports a broader set of metrics computed on the held-out test set, intended for reporting rather than selection. This separation ensures that model ranking is based on resampled estimates while final performance is assessed on data unseen during both training and selection.
\item
  \texttt{plot.fastml(object)} -- generates diagnostic visualizations including metric bar charts, ROC curves, calibration plots, residual diagnostics, and learning curves.
\item
  \texttt{save\_fastml()} / \texttt{load\_model()} -- serialization and deserialization of fastml objects via \texttt{saveRDS}/\texttt{readRDS}.
\end{itemize}

\textbf{Minimal workflow example.} The simplest valid call requires only data and a label:

\begin{Shaded}
\begin{Highlighting}[]
\FunctionTok{library}\NormalTok{(fastml)}
\NormalTok{fm }\OtherTok{\textless{}{-}} \FunctionTok{fastml}\NormalTok{(}\AttributeTok{data =}\NormalTok{ iris, }\AttributeTok{label =} \StringTok{"Species"}\NormalTok{)}
\FunctionTok{summary}\NormalTok{(fm)}
\FunctionTok{predict}\NormalTok{(fm, }\AttributeTok{newdata =}\NormalTok{ new\_obs)}
\end{Highlighting}
\end{Shaded}

\textbf{Explanation and exploration utilities.} \texttt{fastexplain(object,\ method)} provides post hoc model explanations via external packages, supporting 10 methods including DALEX-based variable importance and partial dependence, LIME, ICE, ALE, surrogate trees, interaction analysis, breakdown contributions, counterfactual explanations, fairness diagnostics, and modelStudio dashboards. \texttt{fastexplore(data)} provides pre-modeling exploratory diagnostics (variable summaries, missingness patterns, distributions, correlations, and visualizations) that remain architecturally separate from the training pipeline, as discussed in the Software architecture and security section.

\textbf{Parallel execution.} When \texttt{n\_cores\ \textgreater{}\ 1}, fastml registers a \textbf{doFuture} backend with \texttt{future::multisession} workers. Parallel plans are automatically restored on function exit, and seed handling is configured through \texttt{future.seed} to support deterministic parallel RNG streams. Bootstrap confidence intervals (\texttt{bootstrap\_ci\ =\ TRUE}, default) are computed over \texttt{bootstrap\_samples} (default 500) draws from the evaluation predictions. The package passes CRAN checks and includes a test suite covering core workflows; documentation is available through standard R help pages and a package vignette.

As a rough guide, Case Study B (5-fold CV with three algorithms and tuning grids on N \(\approx\) 8,500) completes in approximately 1.6 minutes on a single core (Intel Core i5-11400 @ 2.60 GHz) using R 4.5. Parallel execution with \texttt{n\_cores\ =\ 4} reduces this to approximately 0.9 minutes. These timings are indicative and will vary with hardware, algorithm complexity, and dataset size.

\subsection{Native survival analysis implementation}\label{sec:native-survival}

\subsubsection{Survival support: learners, outcomes, and resampling}\label{survival-support-learners-outcomes-and-resampling}

\textbf{fastml} supports a set of survival learners that require explicit handling of censoring and, for some engines, specialized prediction objects. In the current release, \textbf{fastml} includes Cox family models: Cox proportional hazards (\texttt{cox\_ph}), penalized Cox regression (\texttt{penalized\_cox}), stratified Cox regression (\texttt{stratified\_cox}), and time varying Cox regression (\texttt{time\_varying\_cox}). It also includes parametric survival models: Accelerated Failure Time regression (\texttt{survreg}), generic parametric survival (\texttt{parametric\_surv}), Royston Parmar flexible parametric survival (\texttt{royston\_parmar}), and a custom piecewise exponential model (\texttt{piecewise\_exp}). Machine learning survival learners include oblique random survival forest survival (\texttt{rand\_forest\_survival}) and XGBoost accelerated failure time (\texttt{xgboost\_aft}). These methods are trained through a shared interface but are dispatched to different fitting backends depending on method and configuration.

\textbf{Outcome representation.}
\textbf{fastml} represents survival outcomes internally using a survival response constructed from follow-up time and event status (i.e., a \texttt{Surv(time,\ status)} object) for standard right-censored data \citep{Therneau:2000_survival_book, Therneau:2024_survival_Rpkg}. For models that require alternative encodings, \textbf{fastml} constructs the appropriate target representation explicitly. In particular, the \texttt{xgboost\_aft} backend uses lower and upper interval bounds for each observation (exact bounds for events; one-sided bounds for right-censoring) rather than a single response value, aligning with the AFT objective function \citep{Barnwal:2020_arXiv}. Basic validation checks ensure that the required outcome fields are present and can be coerced into the expected form, but \textbf{fastml} does not modify censoring patterns (e.g., it does not rebalance events or alter follow-up times).

\textbf{Prediction conventions.}
Survival engines differ in what they naturally return (risk scores, linear predictors, survival curves, time quantiles, or distributional parameters). \textbf{fastml} therefore standardizes predictions only to the extent needed for evaluation: Cox-type learners provide a risk score/linear predictor suitable for concordance-style evaluation \citep{Harrell:1982aa}, parametric learners provide model-based predictions on the scale implied by the fitted distribution, and \texttt{xgboost\_aft} provides predictions aligned with its AFT objective (typically on the log-time scale, with evaluation defined accordingly). When metrics require time-indexed quantities, evaluation uses explicitly supplied time points rather than assuming that every engine can natively return a full survival curve.

\textbf{Fitting under resampling and tuning.}
\textbf{fastml} uses conditional dispatch to choose between workflow-based fitting (via tidymodels abstractions) and direct engine calls. When resampling and/or hyperparameter tuning is requested and a method is supported through a resampling-aware workflow, \textbf{fastml} fits preprocessing and model training within each split so that fold-local preprocessing is preserved. Methods that are not executed through the workflow-based resampling path are fit using their native interfaces and evaluated under the corresponding train/test or resampling configuration supported by the training pipeline. Practically, this means that ``resampling-aware'' fitting is method- and configuration-dependent, and survival resampling is more constrained than for classification/regression tasks.

A minimal survival call illustrates the unified interface. The example below uses \texttt{cox\_ph}, which depends only on the \textbf{survival} package (always available). Additional survival algorithms such as \texttt{penalized\_cox} and \texttt{rand\_forest\_survival} are available when the optional \textbf{censored} package is installed.

\begin{Shaded}
\begin{Highlighting}[]
\FunctionTok{library}\NormalTok{(fastml)}
\FunctionTok{library}\NormalTok{(survival)}
\FunctionTok{library}\NormalTok{(censored)}

\NormalTok{lung\_df }\OtherTok{\textless{}{-}}\NormalTok{ lung[}\FunctionTok{complete.cases}\NormalTok{(lung), ]}
\NormalTok{lung\_df}\SpecialCharTok{$}\NormalTok{status }\OtherTok{\textless{}{-}}\NormalTok{ lung\_df}\SpecialCharTok{$}\NormalTok{status }\SpecialCharTok{{-}} \DecValTok{1}  \CommentTok{\# recode from 1/2 to 0/1}

\NormalTok{fm\_surv }\OtherTok{\textless{}{-}} \FunctionTok{fastml}\NormalTok{(}
  \AttributeTok{data       =}\NormalTok{ lung\_df,}
  \AttributeTok{label      =} \FunctionTok{c}\NormalTok{(}\StringTok{"time"}\NormalTok{, }\StringTok{"status"}\NormalTok{),}
  \AttributeTok{algorithms =} \StringTok{"cox\_ph"}\NormalTok{,}
  \AttributeTok{metric     =} \StringTok{"c\_index"}\NormalTok{,}
  \AttributeTok{seed       =} \DecValTok{2025}
\NormalTok{)}

\FunctionTok{summary}\NormalTok{(fm\_surv, }\AttributeTok{type =} \StringTok{"metrics"}\NormalTok{)}
\end{Highlighting}
\end{Shaded}

\begin{Verbatim}[fontsize=\scriptsize]
## 
## ===== fastml Model Summary =====
## Task: survival 
## Number of Models Trained: 1 
## 
## -- Table 1: Model Selection (Cross-Validation) --
## Note: This table determines the best model.
## 
## --------------------------------------------------------------------- 
## Model    Engine    Harrell C-index (CV mean)  Harrell C-index (CV SD) 
## --------------------------------------------------------------------- 
## cox_ph†  survival  0.6064                     0.1410                  
## --------------------------------------------------------------------- 
## † Selected based on mean Harrell C-index across CV folds
## 
## -- Table 2: Final Evaluation (Test Set) --
## Note: For reporting only; selection was based on CV above.
## 
## ------------------------------------------------------------------------------------------------------------------------ 
## Model   Engine    Harrell C-index  Uno's C-index  Integrated Brier Score  RMST diff (t<=655)  Brier(t=294)  Brier(t=526) 
## ------------------------------------------------------------------------------------------------------------------------ 
## cox_ph  survival  0.660            0.702          0.208                   82.158              0.233         0.238        
## ------------------------------------------------------------------------------------------------------------------------
\end{Verbatim}

\textbf{Explicit non-support and constraints.}
The survival interface is intentionally scoped. Not all survival data structures and evaluation designs are supported in every backend. For example, the \texttt{xgboost\_aft} path supports right-censoring through interval bounds but does not accept counting-process (start-stop) outcomes. More generally, some combinations of survival method, resampling design, and tuning configuration are restricted by the training pipeline (and may be rejected with an error) rather than silently coerced. These constraints are treated as part of the package's design: \textbf{fastml} aims to make supported survival workflows easy to execute correctly, while failing fast when a requested configuration is not implemented or would be ambiguous.

\subsubsection{XGBoost accelerated failure time model with interval bounds}\label{xgboost-accelerated-failure-time-model-with-interval-bounds}

One of the native survival implementations in \textbf{fastml} is the Accelerated Failure Time (AFT) model using XGBoost's survival objective. This path does not rely on a formula interface. Instead, \textbf{fastml} constructs the necessary inputs explicitly and calls the lower-level XGBoost API to ensure correct handling of censoring. The AFT model assumes a parametric form for log-survival time, modeling
\begin{equation}
  \log (T_i) = \eta_i + \varepsilon_i,
\end{equation}
where \(\eta_i\) is a predictor-dependent location parameter and \(\varepsilon_i\) follows a specified error distribution (e.g., Normal, Logistic, or Extreme Value) \citep{Kalbfleisch:2002aa}. To represent censoring correctly, XGBoost requires interval-censored targets rather than single point estimates.

For each observation with observed time \(t_i\) and event indicator \(\delta_i\), \textbf{fastml} computes
\begin{equation}
  \ell_i = \log(t_i),
\end{equation}
and defines the target interval as \([\ell_i, u_i)\) where \(u_i = \ell_i\) for uncensored observations (\(\delta_i = 1\)) and \(u_i = \infty\) for right-censored observations (\(\delta_i = 0\)). These lower and upper bounds are supplied to XGBoost by setting \texttt{label\_lower\_bound} and \texttt{label\_upper\_bound} in the \texttt{xgb.DMatrix}. By doing so, \textbf{fastml} ensures that right-censored observations contribute a one-sided constraint to the loss function, indicating that the event time exceeds the observed value without assuming an exact failure time. This avoids treating censored times as exact regression targets \citep{Barnwal:2020_arXiv}.

The implementation operates directly on log-time and interval bounds and does not support start-stop (counting process) survival data; such outcomes are explicitly rejected for the XGBoost AFT path. \textbf{fastml}'s role here is to translate standard right-censored survival data into the interval-censored representation required by XGBoost's AFT objective, without altering the underlying censoring information or relying on inappropriate regression abstractions.

\subsubsection{Custom piecewise exponential distribution}\label{custom-piecewise-exponential-distribution}

In addition to standard parametric survival models, \textbf{fastml} implements a custom piecewise exponential distribution integrated with the \textbf{flexsurv} framework \citep{Jackson:2016aa}. The goal is to allow flexible baseline hazard shapes while retaining a parametric likelihood and interpretable parameters.

The piecewise exponential model assumes that the hazard function is constant within predefined time intervals \citep{Friedman:1982aa}. Let \(0 < c_1 < c_2 < \cdots < c_K\) denote user-supplied cutpoints. Within each interval \((c_{k-1}, c_k]\), the hazard is assumed to be
\begin{equation}
  h(t) = \lambda_{k},
\end{equation}
where each \(\lambda_k > 0\) is an interval-specific hazard rate. \textbf{fastml} implements this model by defining the hazard, cumulative hazard, density, distribution, and quantile functions algorithmically in a custom distribution family compatible with \textbf{flexsurv}.

User-provided breakpoints are normalized to positive, finite values; \textbf{fastml} does not automatically append an infinite upper bound, nor does it enforce a specific number of intervals. Internally, the custom distribution parameterizes hazards using a baseline log-rate (\texttt{log\_rate}) and a sequence of log hazard ratios (\texttt{log\_ratio\_1}, \texttt{log\_ratio\_2}, \ldots), which are then converted into interval-specific rates \(\lambda_k\). This parameterization is used to ensure positivity and stable estimation while still yielding hazard rates that correspond to piecewise-constant intervals.

The cumulative hazard at time \(t\) is computed as the sum of hazards over all completed intervals plus the partial contribution of the interval containing \(t\). The survival function is then
\begin{equation}
  S(t) = \exp(-H(t)),
\end{equation}
and the density follows directly from the hazard and survival functions. These expressions correspond to the standard piecewise constant hazard formulation, while the implementation itself evaluates them procedurally.

Model fitting is performed via \texttt{flexsurv::flexsurvreg()}, with \textbf{fastml} passing the custom distribution definition and the normalized cutpoints through its internal fitting utilities. This allows the piecewise exponential model to be estimated using maximum likelihood in the same manner as built-in \textbf{flexsurv} distributions, while offering greater flexibility than single-parameter parametric forms such as the exponential or Weibull \citep{Lawless:2003aa}. Each estimated interval hazard rate retains an interpretation as the event rate within a specified time band, making the model useful when hazard rates are expected to vary across follow-up time.

\subsection{Software architecture and security}\label{sec:software-architecture-and-security}

Beyond statistical validity, an AutoML framework must be robust as software. In practice, this means returning results in a form that is easy to inspect, serialize, and reuse, while also reducing the chance that extensibility mechanisms introduce hidden dependencies or unintended side effects \citep{Wilson:2014aa}. In \textbf{fastml}, these goals are addressed through (i) a structured S3 object \citep{Wickham:2019aa} that centralizes fitted models, preprocessing, evaluation outputs, and (optionally) the data partitions used during training and evaluation, and (ii) a set of security and audit utilities that can surface risky patterns in user-supplied recipes and hooks.

The audit and validation utilities are designed to reduce hidden dependencies and common misuse, such as reliance on global environment variables which harms reproducibility \citep{Peng:2011aa}, but they do not provide data-security guarantees; confidentiality depends on what the user chooses to retain and how objects are stored and shared.

\subsubsection{The object model: a self-contained result object}\label{the-object-model-a-self-contained-result-object}

Many R workflows store fitted models, preprocessing artifacts, predictions, and evaluation results as separate objects, or recreate them on demand. \textbf{fastml} instead returns a single S3 object that aggregates the key artifacts produced during training and evaluation. The object is intended to be sufficient for prediction and plotting in a stable/compatible R environment, given the stored fitted model objects and preprocessing specification. However, portability across machines or R installations is not guaranteed for all engines or workflows, because some fitted objects may contain external pointers or compiled components \citep{Eddelbuettel:2013}, serialization behavior can be version-dependent, and prediction may depend on the availability and behavior of upstream packages and recipe steps. Reproducibility also depends on package versions, RNG state, and platform-specific numerical differences \citep{Stodden:2014aa}. In particular, the \textbf{fastml} object:

\begin{itemize}
\tightlist
\item
  stores fitted model objects, organized by algorithm and engine,
\item
  stores the preprocessing object used for prediction,
\item
  stores evaluation artifacts used for reporting (notably holdout/test predictions and performance summaries),
\item
  and may retain the raw and processed training/test partitions used for fitting and evaluation, enabling inspection and post hoc validation. This improves portability and auditability, but also means the object may carry data and is therefore not ``data-free''.
\end{itemize}

The object represents either a resampling-based workflow or a fixed holdout workflow. When resampling is requested and executed, the object records the resampling plan and summary performance outputs associated with that plan. When a single train-test split is used, resampling-related fields are absent or \texttt{NULL}, and reported metrics refer to the holdout/test evaluation.

Table \ref{tab:fastml-object-table} below summarizes common components of a \textbf{fastml} object and the practical guarantees they provide.

\begin{landscape}
\renewcommand{\arraystretch}{1.2}
\begin{table}[!htb]
  \footnotesize
  \centering
  \caption{Core components of a \textbf{fastml} object and associated guarantees.}\vspace{2pt}\label{tab:fastml-object-table}
  \begin{tabular}{L{4cm} L{5.2cm} L{11cm}}
    \toprule
    Slot & Description & Practical guarantee \\
    \midrule
    \texttt{models} & Fitted model objects, indexed by algorithm and engine & Models are stored as fitted objects; downstream \texttt{predict}/\texttt{plot} methods are intended to use stored fits rather than refitting. Reproducibility depends on package versions and environment consistency. \vspace{2pt}\\
    \texttt{preprocessor} & Stored preprocessing object & The stored preprocessor corresponds to the final training split (\texttt{train\_data}), not to per-fold analysis subsets. Fold-local preprocessing occurs inside resampling when that path is used, but the stored final preprocessor is trained on the training split used for the final fit. \vspace{2pt}\\
    \texttt{predictions} & Stored predictions from evaluation & Predictions stored here are the holdout/test predictions produced during evaluation. Training-set predictions and per-fold resampling predictions are not necessarily stored in the same slot by default. \vspace{2pt}\\
    \texttt{performance} & Aggregated performance metrics & Metrics are computed from evaluation outputs (typically holdout/test predictions). If resampling is used, summary performance reflects the resampling results recorded by the resampling layer. \vspace{2pt}\\
    \texttt{metric} & Primary evaluation metric & Defines the main comparison target used in model reporting within the object. \vspace{2pt}\\
    \texttt{best\_model\_name} & Identifier of the default reference model & A presentation/default-selection label; it should not be interpreted as changing how metrics were computed. \vspace{2pt}\\
    \texttt{resampling\_plan} & Resampling specification (e.g., CV folds, repeats) & Encodes how resampling was defined when resampling is executed; used for reporting and consistency. \vspace{2pt}\\
    \texttt{raw\_train\_data, raw\_test\_data} & Data partitions used for fitting/evaluation (when retained) & Enables inspection of the exact splits used. This supports auditability, but the object may include data and therefore is not data-free. \vspace{2pt}\\
    \texttt{processed\_train\_data, processed\_test\_data} & Preprocessed partitions (when retained) & Supports inspection of preprocessing output and helps detect unexpected preprocessing behavior; this is an audit convenience, not a mathematical guarantee of leakage absence. \vspace{2pt}\\
    \texttt{audit} & Audit log entries and a flagged indicator & When \texttt{audit\_mode = TRUE}, the audit structure records security-related warnings/flags. It is not a full training provenance log. \vspace{2pt}\\
    \bottomrule
  \end{tabular}
\end{table}
\end{landscape}

This object contract is used throughout the package. Plotting methods are intended to operate on stored predictions and performance summaries rather than recomputing from scratch. Similarly, \texttt{predict.fastml} applies the stored preprocessor and stored fitted model(s), yielding consistent predictions within a stable R environment. The object is designed to be serializable (e.g., via \texttt{saveRDS}), enabling reuse in downstream workflows.

\subsubsection{Security and audit utilities}\label{security-and-audit-utilities}

AutoML frameworks often support extensibility through custom recipe steps, custom metrics, or user hooks. This extensibility introduces two practical risks: (i) hidden dependencies (for example, reliance on \texttt{.GlobalEnv} or objects defined outside the modeling call), which undermines reproducibility and auditability, and (ii) side effects (for example, reading from files, writing outputs to files, or modifying global objects), which can make automated pipelines fragile.

\textbf{fastml} addresses these risks through two distinct mechanisms that serve different purposes: (a) recipe validation, which is preventive but intentionally narrow, and (b) audit-mode instrumentation, which is broader but observational and incomplete. Neither mechanism should be interpreted as a comprehensive security boundary inside R. Users can bypass many checks via qualified namespace calls (e.g., \texttt{base::}), alternative I/O functions and connections, \texttt{system}, non-local assignment, \texttt{getFromNamespace}, or side effects in compiled code.

\textbf{Recipe validation (preventive, narrow).}
\textbf{fastml} includes a recipe-scanning guard (implemented in \texttt{R/security\_guards.R}) that focuses on a small class of high-impact issues that commonly produce hidden dependencies or fold-external data injection. In particular, it scans recipe steps for patterns suggesting reliance on external environments or global objects (e.g., \texttt{.GlobalEnv}, parent-frame access, or related environment inheritance patterns) and flags steps that appear to embed external data objects (for example, steps inheriting from \texttt{data.frame}). Recipes flagged as depending on external environments or embedded data objects are rejected before model training begins, with an informative message. This guard is not a full static analyzer and it does not attempt to detect file reads/writes or arbitrary unsafe operations inside recipe code.

\textbf{Audit-mode instrumentation (observational, broader but incomplete).}
When \texttt{audit\_mode\ =\ TRUE}, \textbf{fastml} can run parts of the pipeline under an instrumented environment that records and flags selected risk patterns during execution. This includes (i) flagging common forms of global-environment dependence under the instrumented execution path (for example, by detecting attempted \texttt{.GlobalEnv} access and monitoring calls to functions such as \texttt{assign}, \texttt{rm}, and \texttt{get} when they target the global workspace), and (ii) flagging or recording usage of common functions that read from files or write outputs to files (for example, wrappers around frequently used read/write helpers). \textbf{fastml} may also perform lightweight symbol/name scans of function bodies (e.g., via \texttt{all.names}) to flag obvious \texttt{.GlobalEnv} and I/O-related patterns. These checks are best interpreted as audit signals under the instrumented execution path rather than as proof of safety or guarantees of prevention.

\textbf{Integration with the training pipeline.}
Recipe validation is invoked before model training to prevent a narrow class of unsafe recipe definitions. Audit logging is invoked when \texttt{audit\_mode\ =\ TRUE} to emit warnings and record flags. Helper utilities for instrumented execution of arbitrary user hooks may exist, but they are not necessarily used by the default training flow; the audit mechanism should therefore be described as optional and configuration-dependent.

If contrasted with OS-level solutions (e.g., \textbf{RAppArmor} \citep{Ooms:2013aa}), the appropriate framing is that OS-level tools can enforce stronger process constraints, whereas \textbf{fastml} operates within the R runtime and focuses on targeted prevention (recipe validation) plus optional observability (\texttt{audit\_mode}) for common reproducibility risks.

\subsubsection{Exploratory diagnostics and architectural isolation}\label{exploratory-diagnostics-and-architectural-isolation}

Before initiating automated model training, workflows commonly include an exploratory assessment of variable types, missingness patterns, distributions, and data-quality issues. In fastml, this is supported by an optional diagnostic utility, \texttt{fastexplore}, designed to remain separate from the automated resampling and training core.

First, \texttt{fastexplore} is not called internally by \texttt{fastml} and does not participate in model fitting, recipe estimation inside resampling, or evaluation. Second, its default behavior is low side-effect: it returns structured summaries (tables and plots) without writing to disk. Disk output is explicitly opt-in through arguments such as \texttt{save\_results} or \texttt{render\_report}.

In practice, \texttt{fastexplore} produces variable-level summaries (types, unique counts, descriptive statistics), missingness diagnostics (per-variable rates, missing-data pattern visualizations including UpSet plots), distribution checks (normality tests, skewness detection), correlation analysis (pairwise matrices with configurable thresholds for flagging high collinearity), and grouped visualizations (histograms, boxplots, bar charts, heatmaps, and scatterplots). These outputs are intended to inform modeling configuration decisions (for example, whether to impute, transform, or remove features), but they do not automatically alter the training pipeline. Any actions based on diagnostics must be encoded explicitly by the user in the subsequent modeling call.

\subsection{Comparison with existing frameworks}\label{comparison-with-existing-frameworks}

The R ecosystem offers several mature frameworks for machine learning. To contextualize \textbf{fastml}, we compare it against four widely used alternatives: \textbf{caret} \citep{caret_kuhn:2008aa} (legacy unified interface), \textbf{tidymodels} \citep{tidymodels_Kuhn:2020aa} (modular workflow system), \textbf{mlr3} \citep{mlr3_Lang:2019aa} (object-oriented framework with a strong benchmarking culture), and \textbf{h2o} \citep{Fryda:2024_h2o_Rpkg} (an AutoML platform with an internal training/evaluation engine). The comparison below combines (i) architectural differences relevant to leakage risk, survival modeling, and workflow ergonomics, and (ii) external benchmarks conducted for this paper. Implementation details described as properties of \textbf{fastml} refer to its current codebase; statements about other frameworks reflect their documented behavior and typical usage patterns rather than properties established by \textbf{fastml}'s code.

\subsubsection{Feature comparison matrix}\label{feature-comparison-matrix}

Frameworks differ in where they sit on the flexibility-to-automation spectrum and in how explicitly users must assemble preprocessing, resampling, model fitting, and evaluation. \textbf{caret} provides a largely unified training interface; \textbf{tidymodels} emphasizes explicit composition (recipes + model specification + workflow + rsample); \textbf{mlr3} provides an object-oriented task/learner design with strong support for benchmarking; and \textbf{h2o} provides an integrated AutoML system with its own training and evaluation engine. \textbf{fastml} adopts a single-call fa\c{c}ade that orchestrates preprocessing, fitting, and evaluation while exposing a guarded resampling path and diagnostics intended to reduce common workflow errors.

A key comparison dimension is how preprocessing is coupled to evaluation. \textbf{tidymodels} and \textbf{mlr3} are flexible enough to express both leakage-safe and leakage-prone workflows; leakage-resistant evaluation is obtained when preprocessing is estimated on the analysis portion of each resample and then applied to the corresponding assessment data. \textbf{fastml}'s guarded resampling path is designed to re-estimate preprocessing within each resample when that execution mode is used, reducing the risk of preprocessing leakage under supported configurations. This should be interpreted as risk reduction rather than a universal guarantee.

Survival-model support also differs across ecosystems in both breadth and evaluation tooling. \textbf{caret}, \textbf{tidymodels}, and \textbf{mlr3} offer survival modeling through different abstractions and extension packages (e.g., \textbf{censored} package \citep{Hvitfeldt:2025_censored_Rpkg}), while \textbf{fastml} exposes a set of survival methods through a unified interface and dispatches between parsnip-based workflows and native engines depending on whether resampling/tuning is requested, as detailed in the Native survival analysis implementation section. The practical set of methods available to any wrapper framework ultimately depends on engine availability and on which internal execution path is invoked.

Finally, extensibility and ``safety'' features should be interpreted precisely. \textbf{fastml} includes preventive recipe validation plus optional audit-mode instrumentation aimed at surfacing common sources of hidden dependencies and side effects, as described in the Software architecture and security section. These utilities improve auditability under an instrumented execution path, but they do not constitute a security boundary inside R. Table \ref{tab:framework-comparison-table} below highlights differences in design philosophy, safety enforcement, and domain-specific support.

\begin{landscape}
  \begin{table}[!htb]
    \footnotesize
    \centering
    \caption{Architectural and functional comparison of major R machine learning frameworks.}
    \label{tab:framework-comparison-table}
    \begin{tabular}{p{3.2cm}*{5}{L{3.2cm}}}
      \toprule
      Feature & \textbf{caret} \newline (Legacy) & \textbf{tidymodels} \newline (Modular) & \textbf{mlr3} \newline (High-Performance) & \textbf{h2o} \newline (AutoML) & \textbf{fastml} \newline (Guarded Facade) \\
      \midrule
      \multirow{2}{*}{\parbox{3cm}{Interface \newline Philosophy}} & \emph{Unified Wrapper} & \emph{Modular Pipeline} & \emph{Object-Oriented} & \emph{Integrated Engine} & \emph{Guarded Facade}\vspace{2pt}\\
      & Function-based syntax (e.g., \texttt{train()}) & Verbose composition (Recipe + Spec + Workflow) & R6 classes (Tasks, Learners, Graphs) & Server-client AutoML platform & Simplified single-function call (\texttt{fastml()}) \vspace{6pt}\\
      \multirow{2}{*}{\parbox{3cm}{Preprocessing and Leakage Control}} & \emph{Manual} & \emph{Optional/User-Assembled} & \emph{Robust / Graph-Based} & \emph{Automated} & \emph{Enforced Guarded Resampling} \vspace{2pt}\\
      & User must manually prevent leakage; error-prone. & Tools exist (recipes) but isolation is not enforced. & Pipelines define explicit graphs to handle flow. & Internal handling within the H2O engine. & Preprocessing is strictly re-estimated per fold by default. \vspace{6pt}\\
      \multirow{2}{*}{\parbox{3cm}{Survival Analysis Support}} & \emph{Basic} & \emph{Expanding} & \emph{Extensive} & \emph{Standard} & \emph{Native Hybrid} \vspace{2pt}\\
      & Primarily Cox Proportional Hazards & Via \textbf{censored} extension package. & Via \textbf{mlr3} extensions. & Cox AFT models & Unified interface for Penalized Cox + Native XGBoost AFT \& Piecewise Exp. \vspace{6pt}\\
      \multirow{2}{*}{\parbox{3cm}{Extensibility and \newline Safety}} & \emph{Standard Execution} & \emph{Standard Execution} & \emph{Standard Execution} & \emph{Engine-Isolated} & \emph{Sandboxed Execution}\vspace{2pt}\\
      & User code runs directly in R session. & User code runs directly. & User code runs directly. & Runs on Java backend. & Static analysis \& environment masking for user recipes/metrics.\vspace{2pt}\\
      \bottomrule
    \end{tabular}
  \end{table}
\end{landscape}

\section{Results}\label{results}

\subsection{Simulation results}\label{sec:simulation-results}

Across 100 Monte Carlo runs, the leaky workflow achieved a mean ROC AUC of 0.809 (SD: 0.018), whereas the guarded workflow achieved a mean ROC AUC of 0.651 (SD: 0.046).

The paired inflation, defined as Leaky AUC - Guarded AUC within each run, was 0.158 on average, with a 95\% t-based confidence interval of \([0.149, 0.167]\).

These results indicate a large and consistent upward bias in apparent discrimination when site-wise normalization is performed globally prior to grouped validation. The guarded workflow substantially reduces this bias by re-estimating preprocessing within each resample and applying fold-trained transformations only to the corresponding held-out site.

\begin{figure}
\includegraphics[width=1\linewidth,alt={A two-panel figure. Panel A compares the distribution of ROC AUC values for leaky and guarded workflows across simulation runs. Panel B shows paired points and connecting segments, with most guarded values lower than the corresponding leaky values.}]{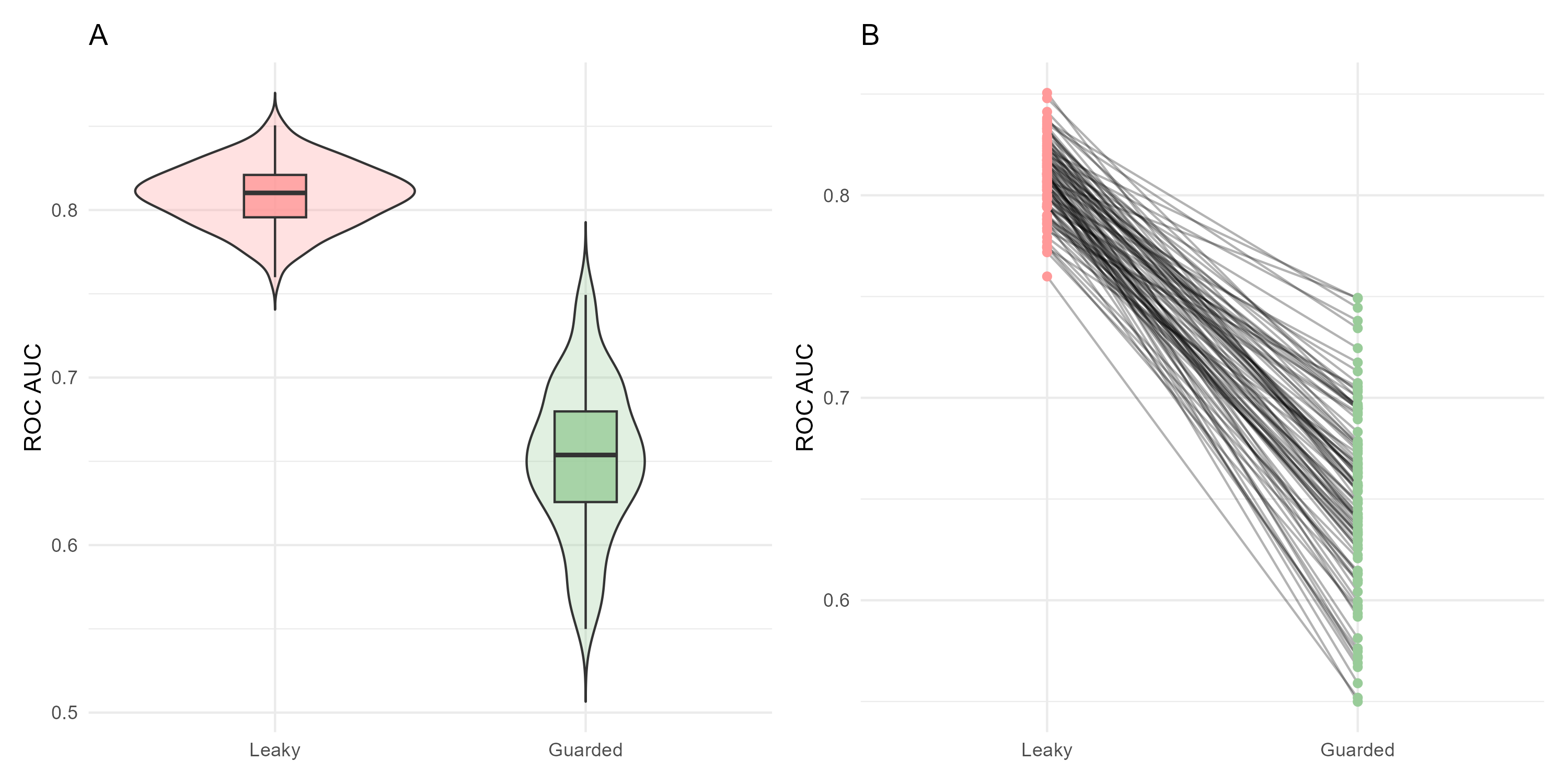} \caption{Preprocessing placement under grouped cross-validation: (A) ROC AUC distributions for leaky versus guarded workflows; (B) paired ROC AUC showing inflation with global preprocessing.}\label{fig:auc-leaky-vs-guarded}
\end{figure}

Figure \ref{fig:auc-leaky-vs-guarded}A shows the distribution of ROC AUC values across runs for the leaky and guarded workflows. Figure \ref{fig:auc-leaky-vs-guarded}B shows the paired within-run change in AUC, highlighting the systematic drop in performance when preprocessing is moved inside the resampling loop.

This experiment demonstrates a concrete and practically relevant failure mode: site-wise normalization performed prior to grouped validation can substantially inflate apparent discrimination, even without access to outcome labels. The paired reductions in ROC AUC show that re-estimating preprocessing within each fold is necessary for valid evaluation when group-level structure is present. In fastml, guarded resampling reduces this class of error by fitting preprocessing and models within resamples when the guarded execution path is used.

\subsection{Benchmarks}\label{benchmarks}

\subsubsection{Benchmark A: usability and code efficiency}\label{benchmark-a-usability-and-code-efficiency}

To quantify differences in user-facing orchestration burden, we implemented a standard clinical classification task using the Pima Indians Diabetes dataset \citep{Smith:1988aa} (N = 768; 8 predictors; missingness in physiologically related variables) in the \textbf{mlbench} package \citep{Leisch:2024_mlbench_Rpkg}. This benchmark is not intended to claim that \textbf{fastml} and \textbf{tidymodels} are definitionally equivalent in preprocessing or modeling defaults. Instead, it compares the amount of explicit pipeline construction typically required to obtain a valid holdout evaluation when preprocessing is fitted on the training partition and then applied to the test partition.

\textbf{Common setup (data splitting).}
We define a single stratified train-test split and supply the same partitions to both approaches so that evaluation is performed on identical holdout data.

\textbf{tidymodels implementation (explicit pipeline).}
A tidymodels workflow typically requires the user to explicitly define preprocessing, assemble a workflow, fit the model, generate predictions, and compute metrics. Because the Pima dataset is entirely numeric, the \texttt{step\_novel}, \texttt{step\_unknown}, and \texttt{step\_dummy} calls below are no-ops for this particular dataset; a minimal, dataset-specific recipe could omit them. We retain them here to illustrate the typical boilerplate a practitioner writes when building a reusable template that must also handle nominal predictors. The code-volume comparison should therefore be read as reflecting a common template-driven workflow rather than the minimal possible recipe for this dataset.

\begin{Shaded}
\begin{Highlighting}[]
\FunctionTok{library}\NormalTok{(tidymodels)}

\NormalTok{rec }\OtherTok{\textless{}{-}} \FunctionTok{recipe}\NormalTok{(diabetes }\SpecialCharTok{\textasciitilde{}}\NormalTok{ ., }\AttributeTok{data =}\NormalTok{ train\_data) }\SpecialCharTok{\%\textgreater{}\%}
  \FunctionTok{step\_impute\_median}\NormalTok{(}\FunctionTok{all\_numeric\_predictors}\NormalTok{()) }\SpecialCharTok{\%\textgreater{}\%}
  \FunctionTok{step\_novel}\NormalTok{(}\FunctionTok{all\_nominal\_predictors}\NormalTok{()) }\SpecialCharTok{\%\textgreater{}\%}
  \FunctionTok{step\_unknown}\NormalTok{(}\FunctionTok{all\_nominal\_predictors}\NormalTok{()) }\SpecialCharTok{\%\textgreater{}\%}
  \FunctionTok{step\_dummy}\NormalTok{(}\FunctionTok{all\_nominal\_predictors}\NormalTok{()) }\SpecialCharTok{\%\textgreater{}\%}
  \FunctionTok{step\_zv}\NormalTok{(}\FunctionTok{all\_predictors}\NormalTok{()) }\SpecialCharTok{\%\textgreater{}\%}
  \FunctionTok{step\_center}\NormalTok{(}\FunctionTok{all\_numeric\_predictors}\NormalTok{()) }\SpecialCharTok{\%\textgreater{}\%}
  \FunctionTok{step\_scale}\NormalTok{(}\FunctionTok{all\_numeric\_predictors}\NormalTok{())}

\NormalTok{log\_spec }\OtherTok{\textless{}{-}} \FunctionTok{logistic\_reg}\NormalTok{() }\SpecialCharTok{\%\textgreater{}\%}
  \FunctionTok{set\_engine}\NormalTok{(}\StringTok{"glm"}\NormalTok{) }\SpecialCharTok{\%\textgreater{}\%}
  \FunctionTok{set\_mode}\NormalTok{(}\StringTok{"classification"}\NormalTok{)}

\NormalTok{wf }\OtherTok{\textless{}{-}} \FunctionTok{workflow}\NormalTok{() }\SpecialCharTok{\%\textgreater{}\%}
  \FunctionTok{add\_recipe}\NormalTok{(rec) }\SpecialCharTok{\%\textgreater{}\%}
  \FunctionTok{add\_model}\NormalTok{(log\_spec)}

\NormalTok{final\_fit }\OtherTok{\textless{}{-}} \FunctionTok{fit}\NormalTok{(wf, }\AttributeTok{data =}\NormalTok{ train\_data)}

\NormalTok{preds }\OtherTok{\textless{}{-}} \FunctionTok{predict}\NormalTok{(final\_fit, }\AttributeTok{new\_data =}\NormalTok{ test\_data) }\SpecialCharTok{\%\textgreater{}\%}
  \FunctionTok{bind\_cols}\NormalTok{(}\FunctionTok{predict}\NormalTok{(final\_fit, }\AttributeTok{new\_data =}\NormalTok{ test\_data, }\AttributeTok{type =} \StringTok{"prob"}\NormalTok{)) }\SpecialCharTok{\%\textgreater{}\%}
  \FunctionTok{bind\_cols}\NormalTok{(test\_data)}

\NormalTok{multi\_metric }\OtherTok{\textless{}{-}} \FunctionTok{metric\_set}\NormalTok{(}
\NormalTok{  accuracy, kap, f\_meas, precision,}
\NormalTok{  sens, spec, roc\_auc}
\NormalTok{)}

\NormalTok{metrics\_tm }\OtherTok{\textless{}{-}} \FunctionTok{multi\_metric}\NormalTok{(}
\NormalTok{  preds,}
  \AttributeTok{truth =}\NormalTok{ diabetes,}
  \AttributeTok{estimate =}\NormalTok{ .pred\_class,}
\NormalTok{  .pred\_pos,}
  \AttributeTok{event\_level =} \StringTok{"second"}
\NormalTok{)}

\FunctionTok{print}\NormalTok{(metrics\_tm)}
\end{Highlighting}
\end{Shaded}

\begin{verbatim}
## # A tibble: 7 x 3
##   .metric   .estimator .estimate
##   <chr>     <chr>          <dbl>
## 1 accuracy  binary         0.776
## 2 kap       binary         0.491
## 3 f_meas    binary         0.656
## 4 precision binary         0.707
## 5 sens      binary         0.612
## 6 spec      binary         0.864
## 7 roc_auc   binary         0.853
\end{verbatim}

\textbf{fastml implementation (single-call interface).}
\textbf{fastml} accepts pre-split data and orchestrates preprocessing, model fitting, prediction, and evaluation within a single call. The specific preprocessing applied depends on the arguments provided and package defaults; therefore, numerical parity with the preceding \textbf{tidymodels} example should only be claimed when preprocessing choices are explicitly matched.

\begin{Shaded}
\begin{Highlighting}[]
\FunctionTok{library}\NormalTok{(fastml)}

\NormalTok{fm }\OtherTok{\textless{}{-}} \FunctionTok{fastml}\NormalTok{(}
  \AttributeTok{train\_data    =}\NormalTok{ train\_data,}
  \AttributeTok{test\_data     =}\NormalTok{ test\_data,}
  \AttributeTok{label         =} \StringTok{"diabetes"}\NormalTok{,}
  \AttributeTok{algorithms    =} \StringTok{"logistic\_reg"}\NormalTok{,}
  \AttributeTok{impute\_method =} \StringTok{"medianImpute"}\NormalTok{,}
  \AttributeTok{event\_class   =} \StringTok{"second"}
\NormalTok{)}

\NormalTok{fm}\SpecialCharTok{$}\NormalTok{performance}\SpecialCharTok{$}\NormalTok{logistic\_reg}
\end{Highlighting}
\end{Shaded}

\begin{verbatim}
## # A tibble: 10 x 6
##    .metric     .estimator .estimate .lower .upper .n_boot
##    <chr>       <chr>          <dbl>  <dbl>  <dbl>   <dbl>
##  1 accuracy    binary        0.776  0.714   0.833     500
##  2 kap         binary        0.491  0.360   0.612     500
##  3 sens        binary        0.612  0.496   0.727     500
##  4 spec        binary        0.864  0.8     0.915     500
##  5 precision   binary        0.707  0.587   0.811     500
##  6 f_meas      binary        0.656  0.561   0.741     500
##  7 roc_auc     binary        0.853  0.799   0.903     500
##  8 logloss     <NA>          0.455  0.382   0.540     500
##  9 brier_score <NA>          0.150  0.122   0.183     500
## 10 ece         <NA>          0.0735 0.0650  0.148     500
\end{verbatim}

Because both implementations use the same model engine (\texttt{glm}), the same training/test partition, and equivalent preprocessing steps, the outputs shown above can be compared directly. When preprocessing choices are matched, the two approaches yield the same holdout metrics, confirming that \textbf{fastml} dispatches to the underlying \textbf{tidymodels} engine without altering the fitted model.

Beyond code volume, the comparison highlights a difference in where correctness effort is concentrated. In a manual \textbf{tidymodels} workflow, the user must explicitly define and correctly sequence each preprocessing step, assemble a workflow object, generate predictions, and compute metrics. Each added step creates an opportunity for omission or mis-ordering. \textbf{fastml} reduces this surface area by bundling common preprocessing, fitting, and evaluation into a single call with safe defaults. This lowers the chance of accidental leakage within the supported pipeline.

\subsubsection{Benchmark B: survival analysis across data scales}\label{benchmark-b-survival-analysis-across-data-scales}

To evaluate survival modeling behavior under different data regimes, we conducted an external benchmark comparing classical baselines with fastml-exposed survival engines across datasets of substantially different sizes. This benchmark is intended to illustrate how a single interface can be used to run multiple survival model classes under a consistent evaluation plan. The benchmark is ``external'' in the sense that all splits, metrics, and summaries are defined by the experimental code in this paper rather than being properties enforced by \textbf{fastml} itself. The complete reproducibility code for this benchmark is provided in the supplementary file \texttt{survival\_benchmark.R}.

\textbf{Datasets.}
We used widely available survival datasets representing distinct scales:

\begin{itemize}
\tightlist
\item
  \textbf{Small}: NCCTG Lung Cancer (\texttt{lung}, \(N = 228\)) \citep{Loprinzi:1994aa}
\item
  \textbf{Medium}: Rotterdam Breast Cancer (\texttt{rotterdam}, \(N = 2,982\)) \citep{Foekens:2000aa}
\item
  \textbf{Large}: Serum Free Light Chain (\texttt{flchain}, \(N = 7,874\)) \citep{Robert:2006aa}
\end{itemize}

\textbf{Evaluation protocol.}
Performance was estimated using repeated 5-fold cross-validation (3 repeats). Folds were stratified by event status to stabilize the censoring/event mix across splits, a common pragmatic choice in finite-sample survival benchmarking. Reported values summarize the distribution of fold-level concordance scores across all resamples.

\textbf{Models compared.}
We evaluated two classical baselines implemented outside \textbf{fastml} using the same resampling splits and metric definitions: Cox proportional hazards (\texttt{survival::coxph}) and Weibull AFT (\texttt{survival::survreg}), alongside fastml-exposed survival methods (e.g., penalized Cox, random-forest survival, and XGBoost AFT) using fixed, pre-specified engine settings. To avoid confounding evaluation with model selection, hyperparameter tuning was not performed in this benchmark; any non-default parameter settings were chosen a priori and held constant within each dataset.

\textbf{Concordance standardization.}
Because some survival engines may return risk scores whose sign convention differs from the standard (higher risk \(=\) shorter survival), raw concordance values below 0.5 are replaced by \(1 - C\) to ensure a consistent directionality across all models. This ``standardized'' C-index always lies in \([0.5, 1]\), where 0.5 indicates no discrimination.

\textbf{Uncertainty summaries.}
For each dataset--model combination, we report the mean concordance across resamples with the corresponding standard deviation, and \(95\%\) confidence intervals computed from the empirical quantiles of resample scores. Because cross-validation fold estimates are not independent, these intervals are descriptive summaries of resampling variability rather than formal inferential confidence intervals \citep{Bengio:2004aa}.

\textbf{Results reporting.}
Table \ref{tab:external-survival-table} below reports mean C-index \(\pm\) SD with \(95\%\) confidence intervals computed over the resamples (5 folds \(\times\) 3 repeats = 15 evaluations). All numeric values in such a table are external experimental results from this study's benchmarking code.

\renewcommand{\arraystretch}{1.3}
\begin{table}[!htb]
  \scriptsize
  \setlength{\tabcolsep}{3pt}
  \centering
  \caption{Mean standardized Harrell\textsuperscript{\textquotesingle}s C-index $\pm$ SD (95\% CI) across data scales from the external benchmark.}\vspace{2pt}\label{tab:external-survival-table}
  \begin{tabular}{L{2cm} *{5}{L{2.2cm}}}
    \toprule
    Dataset & Cox PH \newline (Baseline) & Weibull AFT \newline (Baseline) & \texttt{fastml} \newline (Penalized Cox) & \texttt{fastml} \newline (XGBoost AFT) & \texttt{fastml} \newline (RF Survival) \\
    \midrule
    Lung \newline (N = 228) & 0.631 $\pm$ 0.064 \newline (0.530 to 0.716) & 0.633 $\pm$ 0.062 \newline (0.535 to 0.716) & 0.641 $\pm$ 0.066 \newline (0.528 to 0.739) & 0.618 $\pm$ 0.050 \newline (0.521 to 0.700) & 0.641 $\pm$ 0.063 \newline (0.517 to 0.734) \vspace{2pt}\\
    Rotterdam \newline (N = 2,982) & 0.690 $\pm$ 0.014 \newline (0.669 to 0.713) & 0.690 $\pm$ 0.014 \newline (0.669 to 0.712) & 0.695 $\pm$ 0.015 \newline (0.673 to 0.718) & 0.705 $\pm$ 0.012 \newline (0.690 to 0.724) & 0.712 $\pm$ 0.014 \newline (0.688 to 0.730) \vspace{2pt}\\
    Flchain \newline (N = 7,874) & 0.794 $\pm$ 0.013 \newline (0.775 to 0.817) & 0.794 $\pm$ 0.013 \newline (0.776 to 0.817) & 0.793 $\pm$ 0.012 \newline (0.776 to 0.816) & 0.795 $\pm$ 0.014 \newline (0.778 to 0.819) & 0.799 $\pm$ 0.013 \newline (0.784 to 0.823) \vspace{2pt}\\
    \bottomrule
  \end{tabular}
\end{table}
\renewcommand{\arraystretch}{1.0}

Any narrative interpretation (e.g., ``nonlinear models outperform linear baselines in medium-scale settings'') should be phrased as conclusions from this benchmark under the stated protocol, not as an implication that fastml enforces those outcomes. The main contribution to highlight here is workflow comparability: multiple survival model classes can be evaluated under a single, consistent interface and metric convention, which reduces the cost of comparing methods across regimes, provided the evaluation design (splits, preprocessing placement, and parameter constraints) is defined carefully by the experimenter.

\subsection{Case studies}\label{case-studies}

This section presents two applied examples that mirror common biomedical workflows: (i) comparative diagnostic modeling for binary classification, and (ii) resampling-based regression benchmarking under a fixed tuning budget. The goal is to show how fastml coordinates preprocessing, model fitting, and evaluation within a single interface while keeping key experimental choices explicit (data filtering, random seed, resampling plan, and tuning grid).

All reported metrics (e.g., AUC, RMSE) are empirical outputs of the specific code shown and therefore depend on the declared evaluation design (holdout vs resampling), tuning scope, random seed, and software environment (package versions and available engines).

\subsubsection{Case study A: diagnostic benchmarking (breast cancer)}\label{case-study-a-diagnostic-benchmarking-breast-cancer}

In diagnostic research, it is common to compare a linear baseline with one or more nonlinear learners to evaluate whether additional model flexibility improves discrimination. We illustrate this pattern using the Wisconsin Breast Cancer dataset (\texttt{mlbench::BreastCancer}, N = 569) \citep{Street:1993aa} and three algorithm classes: logistic regression, random forest \citep{Breiman:2001aa}, and gradient boosting \citep{Chen:2016aa_xgboost}. The primary outcome is the ROC AUC on the evaluation data generated under the specified workflow.

The dataset includes an identifier column (Id), which is not a meaningful predictor and should be excluded to avoid distorted performance estimates. The code below removes Id, ensures the outcome (Class) is observed, and uses complete-case analysis for simplicity. When missingness is substantial or plausibly informative, missing-data handling should be modeled explicitly (e.g., via an imputation recipe) rather than addressed by dropping incomplete rows \citep{Little:2019aa}.

\begin{Shaded}
\begin{Highlighting}[]
\FunctionTok{library}\NormalTok{(fastml)}
\FunctionTok{library}\NormalTok{(mlbench)}
\FunctionTok{library}\NormalTok{(dplyr)}

\FunctionTok{set.seed}\NormalTok{(}\DecValTok{2025}\NormalTok{)}

\FunctionTok{data}\NormalTok{(}\StringTok{"BreastCancer"}\NormalTok{)}

\NormalTok{bc }\OtherTok{\textless{}{-}}\NormalTok{ BreastCancer }\SpecialCharTok{\%\textgreater{}\%}
  \FunctionTok{select}\NormalTok{(}\SpecialCharTok{{-}}\NormalTok{Id) }\SpecialCharTok{\%\textgreater{}\%}           \CommentTok{\# exclude identifier}
  \FunctionTok{filter}\NormalTok{(}\SpecialCharTok{!}\FunctionTok{is.na}\NormalTok{(Class)) }\SpecialCharTok{\%\textgreater{}\%} \CommentTok{\# ensure outcome observed}
  \FunctionTok{na.omit}\NormalTok{()                 }\CommentTok{\# simplify this case study: complete{-}case analysis}

\NormalTok{fm\_bc }\OtherTok{\textless{}{-}} \FunctionTok{fastml}\NormalTok{(}
  \AttributeTok{data          =}\NormalTok{ bc,}
  \AttributeTok{label         =} \StringTok{"Class"}\NormalTok{,}
  \AttributeTok{algorithms    =} \FunctionTok{c}\NormalTok{(}\StringTok{"logistic\_reg"}\NormalTok{, }\StringTok{"rand\_forest"}\NormalTok{, }\StringTok{"xgboost"}\NormalTok{),}
  \AttributeTok{metric        =} \StringTok{"roc\_auc"}\NormalTok{,}
  \AttributeTok{event\_class   =} \StringTok{"second"}
\NormalTok{)}
\end{Highlighting}
\end{Shaded}

\begin{Verbatim}[fontsize=\scriptsize]
## 
## ===== fastml Model Summary =====
## Task: classification 
## Number of Models Trained: 3 
## 
## -- Table 1: Model Selection (Cross-Validation) --
## Note: This table determines the best model.
## 
## --------------------------------------------------------- 
## Model         Engine   ROC AUC (CV mean)  ROC AUC (CV SD) 
## --------------------------------------------------------- 
## rand_forest†  ranger   0.9906             0.0132          
## xgboost       xgboost  0.9847             0.0158          
## logistic_reg  glm      0.9230             0.0217          
## --------------------------------------------------------- 
## † Selected based on mean ROC AUC across CV folds
## 
## -- Table 2: Final Evaluation (Test Set) --
## Note: For reporting only; selection was based on CV above.
## 
## --------------------------------------------------------------------------------------------------------------------------- 
## Model         Engine   Accuracy  F1 Score  Kappa  Precision  Sensitivity  Specificity  ROC AUC  Logloss  Brier Score  ECE   
## --------------------------------------------------------------------------------------------------------------------------- 
## rand_forest   ranger   0.971     0.960     0.937  0.923      1.000        0.955        0.996    0.102    0.026        0.052 
## xgboost       xgboost  0.956     0.938     0.904  0.938      0.938        0.966        0.991    0.268    0.066        0.179 
## logistic_reg  glm      0.934     0.903     0.854  0.933      0.875        0.966        0.965    1.079    0.067        0.071 
## ---------------------------------------------------------------------------------------------------------------------------
\end{Verbatim}

\begin{Shaded}
\begin{Highlighting}[]
\FunctionTok{plot}\NormalTok{(fm\_bc, }\AttributeTok{type =} \StringTok{"roc"}\NormalTok{)}
\end{Highlighting}
\end{Shaded}

\begin{center}\includegraphics[width=0.8\linewidth]{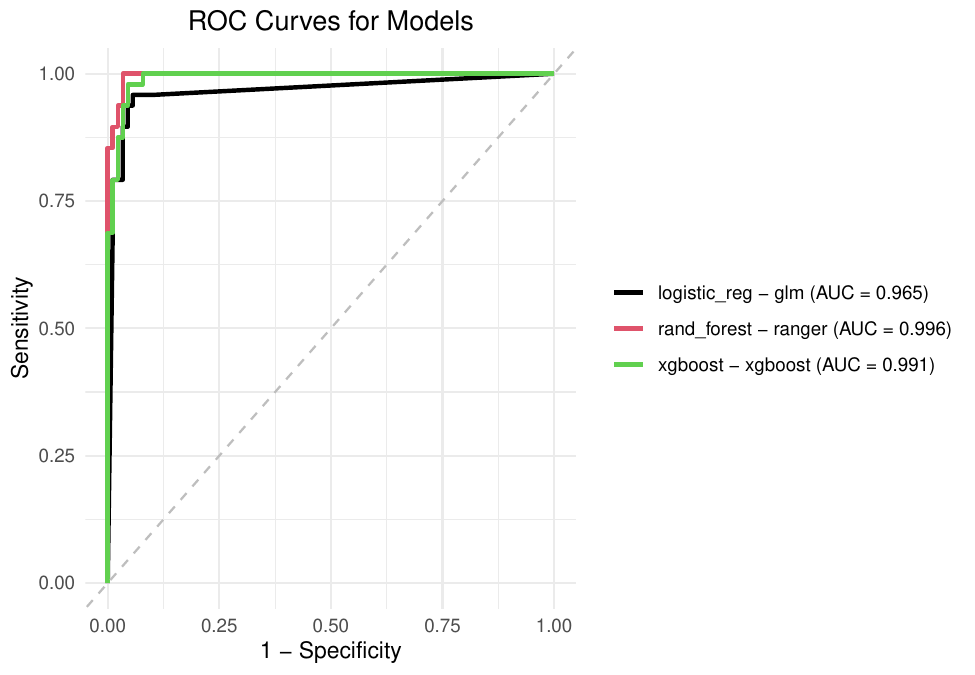} \end{center}

The summary output follows the two-table design described in the Overview. Table 1 ranks the three algorithms by mean ROC AUC across cross-validation folds and selects the best model (marked with \texttt{†}). Table 2 then reports a richer set of held-out test metrics -- accuracy, F1, kappa, precision, sensitivity, specificity, ROC AUC, log-loss, Brier score, and expected calibration error -- for all fitted models. Because selection is based on resampled performance rather than test-set metrics, the test-set evaluation provides an unbiased summary for reporting. The ROC visualization complements these tables by showing sensitivity--specificity trade-offs across classification thresholds.

Model explanation is often desirable in diagnostic applications. \textbf{fastml} provides \texttt{fastexplain} for post hoc explanations via external explainability tooling. The availability of specific explanation artifacts (e.g., permutation importance, SHAP-style outputs) depends on installed explainer packages and on model/engine compatibility \citep{Lundberg:2017aa}; therefore, explanation figures should be reported only when they are produced in the paper's software environment.

\begin{Shaded}
\begin{Highlighting}[]
\CommentTok{\# Optional: explanation outputs depend on installed explainers}
\CommentTok{\#           and model compatibility}
\FunctionTok{fastexplain}\NormalTok{(fm\_bc, }\AttributeTok{method =} \StringTok{"dalex"}\NormalTok{)}
\end{Highlighting}
\end{Shaded}

\begin{verbatim}
## 
## === DALEX Variable Importance (with Boxplots) ===
\end{verbatim}

\begin{center}\includegraphics[width=0.8\linewidth]{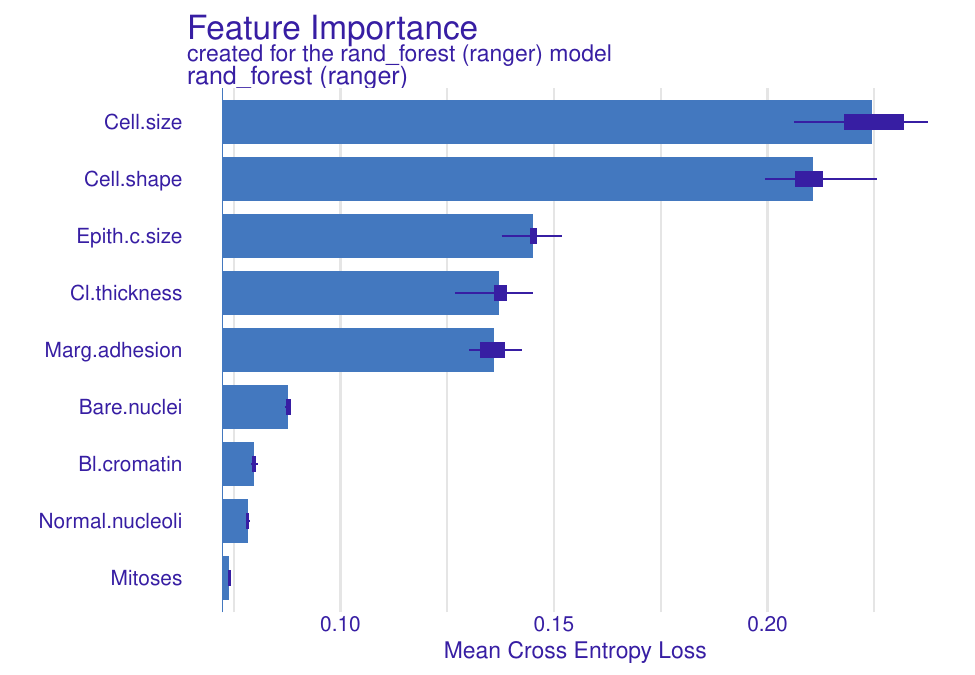} \end{center}

\begin{verbatim}
## 
## === DALEX Shapley Values (SHAP) ===
\end{verbatim}

\begin{center}\includegraphics[width=0.8\linewidth]{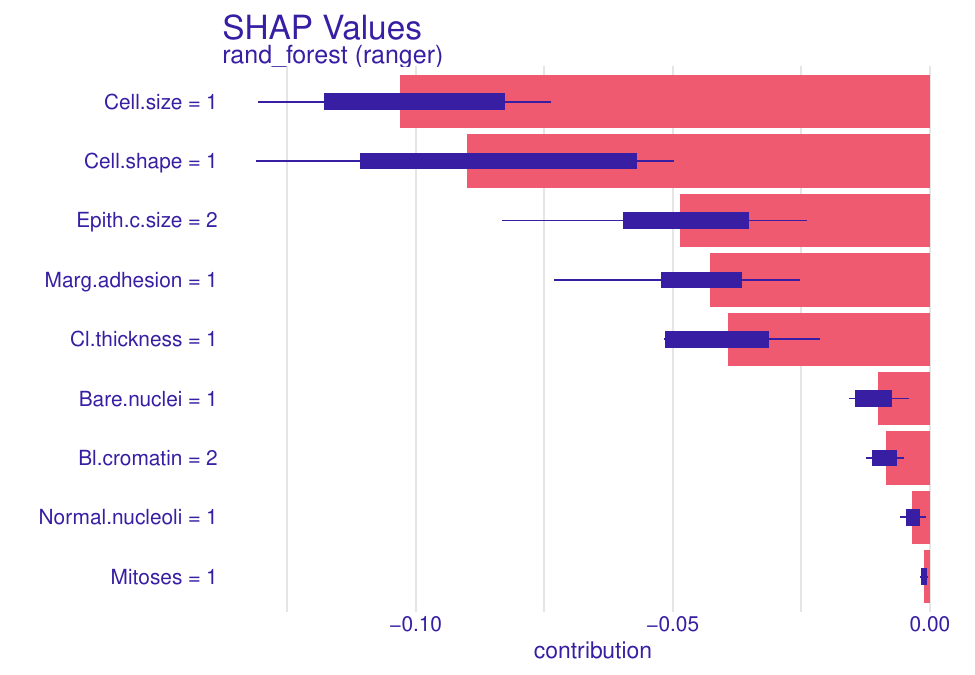} \end{center}

\subsubsection{Case study B: mixed-type clinical regression (hypertension)}\label{case-study-b-mixed-type-clinical-regression-hypertension}

The second example demonstrates resampling-based regression benchmarking in a larger, mixed-type clinical dataset. Using NHANES data (N \(\approx\) 8,500 after filtering to observed outcomes) \citep{Pruim:2025_NHANES_aa}, the objective is to predict systolic blood pressure (\texttt{BPSysAve}) from demographic, lifestyle, and laboratory variables. We compare three model families: elastic net (regularized linear model) \citep{Zou:2005aa}, random forest (bagging ensemble), and LightGBM (gradient boosting) \citep{Ke:2017aa}. Hyperparameters are tuned using a predefined grid under 5-fold cross-validation.

This design reflects a practical goal: evaluate whether a linear model is sufficient, or whether nonlinear learners provide measurable gains under a fixed and transparent tuning budget. Engine availability is an external requirement (e.g., LightGBM via the \textbf{bonsai} package must be installed and accessible), and performance will vary with the tuning grid; thus, results should be presented as outcomes of the specified benchmarking design rather than as general statements about the algorithms.

\begin{Shaded}
\begin{Highlighting}[]
\FunctionTok{library}\NormalTok{(fastml)}
\FunctionTok{library}\NormalTok{(bonsai)}
\FunctionTok{library}\NormalTok{(NHANES)}
\FunctionTok{library}\NormalTok{(dplyr)}

\FunctionTok{set.seed}\NormalTok{(}\DecValTok{2025}\NormalTok{)}

\FunctionTok{data}\NormalTok{(}\StringTok{"NHANES"}\NormalTok{)}

\NormalTok{df }\OtherTok{\textless{}{-}}\NormalTok{ NHANES }\SpecialCharTok{\%\textgreater{}\%}
  \FunctionTok{select}\NormalTok{(}
\NormalTok{    BPSysAve,}
\NormalTok{    Age, Gender, BMI, Race1, Education, Poverty,}
\NormalTok{    SleepHrsNight, PhysActive, Smoke100, TotChol, Diabetes}
\NormalTok{  ) }\SpecialCharTok{\%\textgreater{}\%}
  \FunctionTok{filter}\NormalTok{(}\SpecialCharTok{!}\FunctionTok{is.na}\NormalTok{(BPSysAve))}

\NormalTok{fm\_bp }\OtherTok{\textless{}{-}} \FunctionTok{fastml}\NormalTok{(}
  \AttributeTok{data              =}\NormalTok{ df,}
  \AttributeTok{label             =} \StringTok{"BPSysAve"}\NormalTok{,}
  \AttributeTok{algorithms        =} \FunctionTok{c}\NormalTok{(}\StringTok{"elastic\_net"}\NormalTok{, }\StringTok{"rand\_forest"}\NormalTok{, }\StringTok{"lightgbm"}\NormalTok{),}
  \AttributeTok{metric            =} \StringTok{"rmse"}\NormalTok{,}
  \AttributeTok{impute\_method     =} \StringTok{"medianImpute"}\NormalTok{,}
  \AttributeTok{resampling\_method =} \StringTok{"cv"}\NormalTok{,}
  \AttributeTok{folds             =} \DecValTok{5}\NormalTok{,}
  \AttributeTok{tune\_params       =} \FunctionTok{list}\NormalTok{(}
    \AttributeTok{elastic\_net =} \FunctionTok{list}\NormalTok{(}
      \AttributeTok{glmnet =} \FunctionTok{list}\NormalTok{(}
        \AttributeTok{penalty =} \FunctionTok{c}\NormalTok{(}\FloatTok{0.01}\NormalTok{, }\FloatTok{0.1}\NormalTok{),}
        \AttributeTok{mixture =} \FunctionTok{c}\NormalTok{(}\FloatTok{0.0}\NormalTok{, }\FloatTok{0.5}\NormalTok{, }\FloatTok{1.0}\NormalTok{)  }\CommentTok{\# ridge, elastic net, lasso}
\NormalTok{      )}
\NormalTok{    ),}
    \AttributeTok{rand\_forest =} \FunctionTok{list}\NormalTok{(}
      \AttributeTok{ranger =} \FunctionTok{list}\NormalTok{(}
        \AttributeTok{mtry  =} \FunctionTok{c}\NormalTok{(}\DecValTok{2}\NormalTok{, }\DecValTok{3}\NormalTok{, }\DecValTok{5}\NormalTok{),}
        \AttributeTok{min\_n =} \FunctionTok{c}\NormalTok{(}\DecValTok{20}\NormalTok{, }\DecValTok{50}\NormalTok{),}
        \AttributeTok{trees =} \FunctionTok{c}\NormalTok{(}\DecValTok{100}\NormalTok{)}
\NormalTok{      )}
\NormalTok{    ),}
    \AttributeTok{lightgbm =} \FunctionTok{list}\NormalTok{(}
      \AttributeTok{lightgbm =} \FunctionTok{list}\NormalTok{(}
        \AttributeTok{tree\_depth    =} \FunctionTok{c}\NormalTok{(}\DecValTok{20}\NormalTok{, }\DecValTok{31}\NormalTok{),}
        \AttributeTok{learn\_rate    =} \FunctionTok{c}\NormalTok{(}\FloatTok{0.05}\NormalTok{, }\FloatTok{0.1}\NormalTok{)}
\NormalTok{      )}
\NormalTok{    )}
\NormalTok{  )}
\NormalTok{)}
\end{Highlighting}
\end{Shaded}

\begin{verbatim}
## 
## ===== fastml Model Summary =====
## Task: regression 
## Number of Models Trained: 3 
## 
## -- Table 1: Model Selection (Cross-Validation) --
## Note: This table determines the best model.
## 
## ---------------------------------------------------- 
## Model         Engine    RMSE (CV mean)  RMSE (CV SD) 
## ---------------------------------------------------- 
## rand_forest†  ranger    12.6280         0.3131       
## elastic_net   glmnet    14.3680         0.4643       
## lightgbm      lightgbm  23.6483         0.7209       
## ---------------------------------------------------- 
## † Selected based on mean RMSE across CV folds
## 
## -- Table 2: Final Evaluation (Test Set) --
## Note: For reporting only; selection was based on CV above.
## 
## ------------------------------------------------ 
## Model        Engine    RMSE    R-squared  MAE    
## ------------------------------------------------ 
## rand_forest  ranger    12.053  0.523      8.823  
## elastic_net  glmnet    14.152  0.329      10.483 
## lightgbm     lightgbm  23.058  0.193      15.851 
## ------------------------------------------------ 
## 
## Best Model hyperparameters:
## 
## Model: rand_forest (ranger) 
##   mtry: 5
##   trees: 100
##   min_n: 20
\end{verbatim}

The resulting summary can be used to populate a regression comparison table, reporting RMSE and any additional metrics computed under the same resampling and tuning design. Any conclusions about which model ``wins'' should be tied to the declared tuning grid and evaluation plan: in particular, boosting methods can be sensitive to the search space, and a small grid may not reflect their best attainable performance.

These case studies demonstrate two applied patterns that recur in biomedical modeling: (i) comparative benchmarking across model families for diagnostic classification, and (ii) resampling-based regression benchmarking with explicit tuning budgets. The examples are intentionally written so that preprocessing decisions (e.g., exclusion of identifier fields), evaluation design (seed, resampling plan), and tuning scope are visible and reportable.

\section{Conclusion}\label{conclusion}

The R ecosystem provides increasingly capable machine-learning tooling, but correct evaluation still depends on how users structure preprocessing, resampling, and model selection. A common failure mode is leakage introduced by global data-dependent operations (e.g., imputation, scaling, feature construction) performed before resampling. Such issues can bias performance estimates even when outcome labels are never explicitly used \citep{Kapoor:2023aa}.

\textbf{fastml} is motivated by reducing the practical risk of these errors through a constrained interface and guarded evaluation paths. When \textbf{fastml} executes its guarded resampling path, workflows are fitted within resamples so that data-dependent preprocessing is estimated on analysis splits and then applied to the corresponding assessment splits. This design can substantially reduce leakage risk in typical cross-validation setups.

\textbf{fastml} also extends the same single-call workflow style to survival modeling. In addition to dispatching standard survival learners through established engines, \textbf{fastml} includes native implementations for selected survival methods (e.g., XGBoost AFT with interval bounds and a piecewise-exponential model implemented through flexsurv utilities). This enables side-by-side benchmarking of multiple survival model classes under one interface, while keeping the evaluation design (holdout vs.~resampling) explicit in the analysis.

With respect to operational safety, \textbf{fastml} includes audit utilities intended to surface risky patterns in user-defined functions and recipe steps when \texttt{audit\_mode\ =\ TRUE}. In the current implementation, recipe validation and audit logging can flag \texttt{.GlobalEnv}-style dependencies and I/O-related symbol usage. The audit layer is best characterized as warning/logging infrastructure rather than a hard enforcement mechanism that blocks all unsafe operations.

The empirical results presented in this paper (performance comparisons, model rankings, and usability examples) should be interpreted as outcomes of the experimental designs we executed, defined by specific datasets, preprocessing choices, tuning grids, random seeds, and resampling plans, rather than as properties guaranteed by the implementation. Likewise, any claims about reduced boilerplate relative to alternative frameworks should be treated as qualitative observations or externally measured comparisons, not as invariant guarantees.

\textbf{Limitations.}
Several limitations should be noted. First, guarded resampling protects within the pipeline but not against upstream leakage (e.g., preprocessed inputs supplied by the user). Second, audit mode is observational rather than preventive for all patterns; it flags common risks but cannot intercept arbitrary unsafe operations. Third, survival resampling support is more constrained than for classification and regression tasks, with some method--resampling--tuning combinations explicitly unsupported. Fourth, portability of saved fastml objects depends on R and package versions, since fitted model objects may contain external pointers or version-dependent components. Finally, \textbf{fastml} does not currently support model stacking or ensemble combination; this is noted as a direction for future work.

Future work will focus on expanding the set of supported workflows while preserving evaluation clarity. Two concrete directions are (i) model combination strategies (e.g., stacking/super learning) implemented in a way that keeps resampling and leakage considerations explicit \citep{vanDerLaan:2007aa}, and (ii) improved deployment ergonomics (e.g., generating standardized prediction wrappers and deployment-ready artifacts). These are planned extensions rather than current capabilities and will require careful design to maintain the package's emphasis on transparent evaluation and auditable outputs.

\renewcommand\refname{References}

\end{document}